\documentclass[preprint,showpacs,pre,10pt]{revtex4-1}
\usepackage{amsfonts}
\usepackage{amssymb}
\usepackage{amsmath}
\usepackage{graphicx}
\usepackage{enumerate}
\def\zetant#1{\mathchoice
   {\XXint\displaystyle\textstyle{#1}}%
   {\XXint\textstyle\scriptstyle{#1}}%
   {\XXint\scriptstyle\scriptscriptstyle{#1}}%
   {\XXint\scriptscriptstyle\scriptscriptstyle{#1}}%
   \!\int}
\def\XXint#1#2#3{{\setbox0=\hbox{$#1{#2#3}{\int}$}
     \vcenter{\hbox{$#2#3$}}\kern-.5\wd0}}

\def\dashint{\zetant-}

\begin{document}
\title{Scattering on two Aharonov-Bohm vortices}
\author{E. Bogomolny}
\affiliation{Univ. Paris-Sud, CNRS, LPTMS, UMR8626, F-91405, Orsay, France}

\date{\today}
\begin{abstract}
The problem of two Aharonov-Bohm (AB) vortices for the Helmholtz equation is examined in detail. It is demonstrated that the method proposed in [J. M. Myers, J. Math. Phys. \textbf{6}, 1839 (1963)], Ref.~\cite{myers}, for diffraction on a slit  can be generalized to get an explicit solution for AB vortices. Due to singular nature of AB interaction the Green function and the scattering amplitude for two AB vortices obey a series of partial differential equations. Coefficients entering these equations, in their turn, fulfill ordinary non-linear differential equations whose solutions can be obtained from  a solution of the Painlev\'e V (or III) equation. The asymptotics of necessary functions for very large and very small distances between two vortices are calculated explicitly.  Taken together, it means that the problem of two  AB vortices is integrable.      
\end{abstract}
\maketitle

\section{Introduction}

The Aharonov-Bohm (AB) effect \cite{first}, \cite{AB} is one of the most striking distinguishes  between quantum and classical words. In a nutshell, it states that a quantum particle  feels electro-magnetic potentials even though no classical forces exist. Its first description can be traced to the paper of  Ehrenberg and Siday \cite{first}, but it is only after the seminal work of Aharonov and Bohm \cite{AB} that this subject attracts a wide attention.  The success of that paper can be attributed to the fact that in addition to a general discussion of the phenomenon the authors presented a clear-cut analytic calculation of physical scattering on one singular AB vortex thus validating common arguments.     

Today there exists a huge literature about this effect but, surprisingly, analytically results are rare.   In  \cite{stovicek} a  diagrammatic-like series for the amplitude of scattering on a few AB vortices had been proposed but for real energy it is similar to  a formal multiple scattering expansion and hardly can be used for calculations.    
 
The purpose of this paper is to investigate the problem of scattering on two AB vortices. The principal  result is that this problem  
is integrable and the calculations of the Green function and the scattering amplitude can be reduced to a solution of a series of differential equations whose lowest level includes the Painlev\'e V (or III) equation. The method used in derivation of these results is a generalization of the one proposed  in Ref.~\cite{myers} where the diffraction on a finite slit has been treated. It is based on the point-like nature of the AB potential which permits to fix solutions by fixing its behavior near vortex positions. It means that only a few constants uniquely determine the full solution. Using different transformations commuting with the Laplacian leads to a system of equations for these constants.  Besides equations, it is necessary to know the values of different quantities at small and/or large distances between vortices. For large vortex separation it can be done by perturbation series and for small distances between vortices it is achieved by the using Riemann-Hilbert method. 

The plan of the paper is the following. Section~\ref{general_considerations} is devoted to a general discussion of the problem of scattering on AB vortices. Special attention in this Section is focused on the uniqueness of the solution  and, in particular, on the fact that any solution obeying all boundary and radiation  conditions but without in-coming incident waves is identically zero.  In Sections~\ref{green_function}-\ref{scattering_amplitude} it is demonstrated how the arguments of Ref.~\cite{myers} can be generalized to the case of two vortices. First of all,  a set of auxiliary functions independent of incident fields with prescribed singularities at vortex positions are introduced. These functions  are  analogues of the Hankel functions for one-vortex problem and play a prominent role in what follows. The main idea of Ref.~\cite{myers} is that there exists a group of differential operators which commute with the Lagrangian and cancel the incident field.  Transformed solution is non-zero as the action of these group operators change boundary conditions near vortices. But these changes can be compensated by a suitable linear combination of new functions. In this manner one gets a set of equations for unknown functions. Calculation of group commutators done in Section~\ref{determination_constants} permits to find equations for all necessary functions.  In Section~\ref{green_function} the Green function is discussed and in Section~\ref{scattering_amplitude} this procedure is done for the scattering amplitude. To use the obtained equations one needs to find the asymptotics of correct solutions at small and/or at  large separation between vortices. This is achieved in Section~\ref{small_distance} where  explicit forms of the solution when the distance between vortices tends to zero and to infinity are obtained. Section~\ref{conclusion} is a summary of the obtained results. The relation of these results to the theory of holonomic quantum field \cite{swj} is in short discussed here. 
Appendix~\ref{uniqueness} is devoted to the proof of the uniqueness of the solution and to derivation of the reciprocity relation for the scattering on two AB vortices. In Appendix~\ref{one_vortex} properties of one-vortex solution are briefly discussed.  

To diminish the paper size only the main steps of derivations are presented and details are often omitted. 


\section{General considerations}\label{general_considerations}

The AB vector potential, $A_{\mu}$, is a pure gauge potential, $ A_{\mu}=\partial_{\mu}\phi$, and can be removed by a gauge transformation. Nevertheless, the existence of AB vortices manifests in  non-zero circulation along any closed contour encircling only  vortex $j$
\begin{equation}
\oint A_{\mu}\mathrm{d}x_{\mu}=-2\pi \alpha_j,
\end{equation}    
where $\alpha_j$ is the magnetic flux associated  with the vortex (we assume that $\alpha\neq $ integer). 

\begin{figure}
\begin{center}
\includegraphics[width=.25\linewidth]{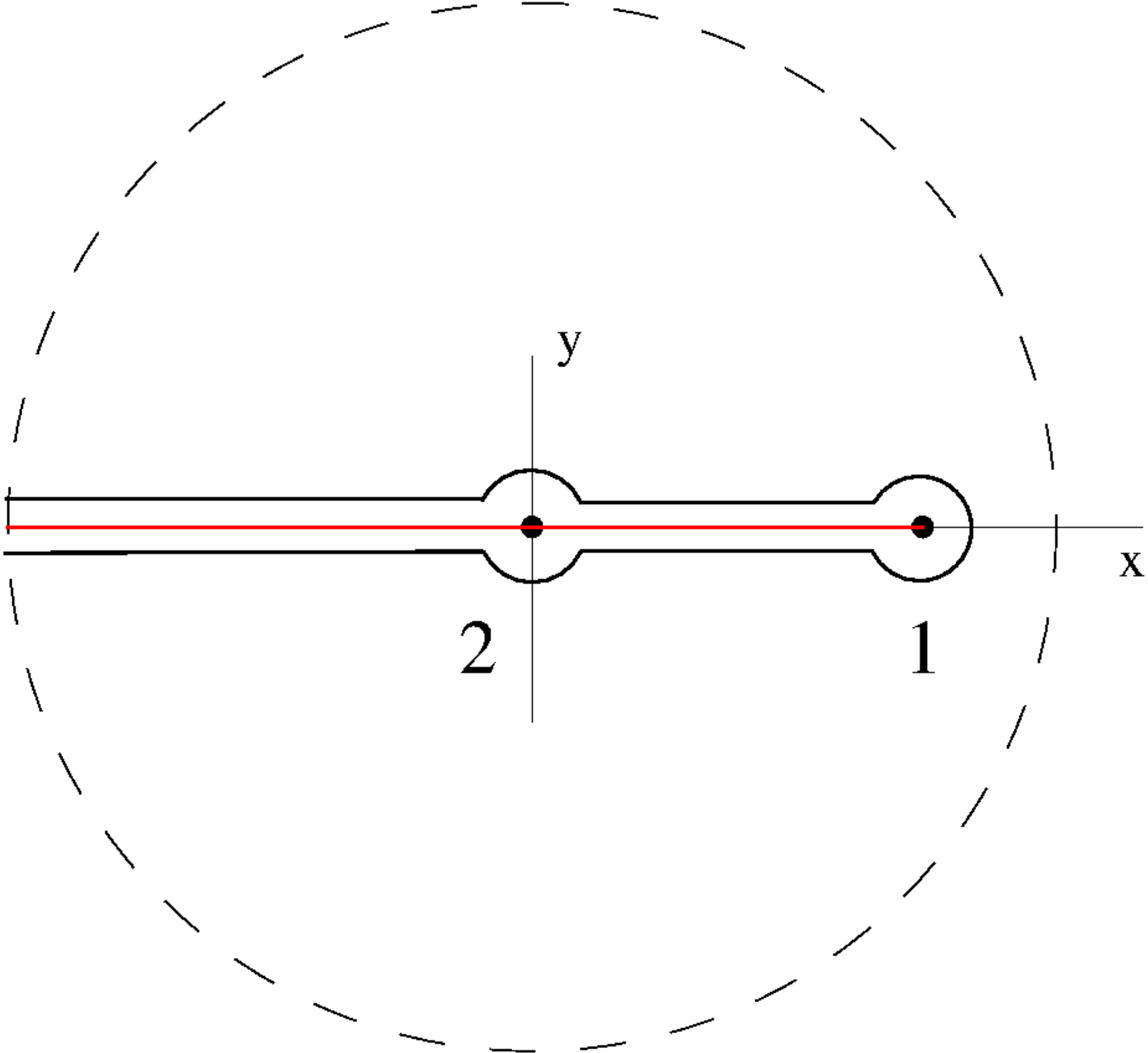}
\end{center}
\caption{Cuts for the two-vortex problem (red lines). Black solid lines are contours  encircling the  vortices.  Dashed circle indicates large radius contour around all vortices. }
\label{contour}
\end{figure}

The existence of a non-zero circulation implies that after potential removing from each vortex  emanates a line of phase discontinuity  (the cut) denoted  $\mathcal{C}_j$ such that the function and its normal derivative on the both sides of the cut differ by a phase
\begin{equation}
\Psi_+(x,0)=\mathrm{e}^{2\pi \mathrm{i}\alpha_j}\Psi_-(x,0),\qquad 
\partial_y \Psi_+(x,0)=\mathrm{e}^{2\pi \mathrm{i}\alpha_j}\partial_y \Psi_-(x,0),\qquad x\in \mathcal{C}_j\ .
\label{boundary}
\end{equation}
Here $x$ and $y$  are coordinate respectively along and perpendicular to the cut. Each cut has two different sides which can be connected by a contour encircling  one or more vortices.  The cuts can be chosen arbitrarily and wave functions with different cuts are  gauge equivalent. In Fig.~\ref{contour} a convenient choice of cuts for two AB vortices used throughout the paper  is sketched.         
In this case the two cuts coincide along the negative $x$-axis and boundary values of wave function and its $y$-derivative on the cut are related as follows
\begin{equation}
\Psi_+(x,0)=\mathrm{e}^{2\pi \mathrm{i}\chi(x) }\Psi_-(x,0),\qquad 
\partial_y \Psi_+(x,0)=\mathrm{e}^{2\pi \mathrm{i} \chi(x)}\partial_y \Psi_-(x,0),
\label{boundary_uniform}
\end{equation}
 where piece-wise constant function $\chi(x)$ is 
\begin{equation}
\chi(x)=\left \{\begin{array}{rr}\alpha_1+\alpha_2 ,& x<0\\ \alpha_1,& 0<x<L\\0,&x>0\end{array}\right .\ , 
\label{chi}
\end{equation}  
and subscripts $(\pm)$ corresponds to limit $y\to 0$ from positive and  negative values of $y$ respectively. 
The problem of one vortex has been solved in \cite{AB} (and is shortly reviewed in Appendix~\ref{one_vortex}).
\vspace{.4cm} 

As any problem of diffraction, the scattering on the AB vortices corresponds to finding  a wave function $\Psi(\vec{x}\,)$ with the following properties:
\begin{enumerate}[(a)]
\item  $\Psi(\vec{x}\,)$ is the sum of an incident wave $\Psi^{\mathrm{inc}}(\vec{x})$ which includes all in-coming waves 
and a reflected out-going wave $\Psi^{\mathrm{ref}}(\vec{x}\,)$,
\begin{equation}
\Psi(\vec{x}\, )=\Psi^{\mathrm{inc}}(\vec{x}\, )+\Psi^{\mathrm{ref}}(\vec{x}\, )\ .
\end{equation}
The choice of the incident wave is dictated by the problem considered. When one is interested in  the Green function,  the incident wave is the Green function of the free Helmholtz equation
\begin{equation}
\Psi^{\mathrm{inc}}(\vec{x}\, )=\frac{1}{4\mathrm{i}}H_0^{(1)}(k|\vec{x}-\vec{x}^{\, \prime}|)
\label{free_hankel}
\end{equation} 
with $\vec{x}^{\, \prime}$ being the source point. For the problem of the plane wave scattering
\begin{equation}
\Psi^{\mathrm{inc}}(\vec{x}\, )=\mathrm{e}^{\mathrm{i}k r \cos (\theta-\phi)},
\label{plane_wave}
\end{equation} 
where $r$ and $\theta$ are polar coordinates of point $\vec{x}$ and $\phi$ is the angle of incidence. 

\item  The reflected field  obeys the Helmholtz equation
\begin{equation}
(\partial_x^2+\partial_y^2 +k^2)\Psi^{\mathrm{ref}}(\vec{x}\, )=0
\label{helmholtz}
\end{equation}    
everywhere in the plane $(x,y)$ except the cuts. 

\item  At the both sides of the cuts the full wave function and its normal derivative are related as in  Eq.~\eqref{boundary} (or for the cuts in Fig.~\ref{contour} as in Eq.~\eqref{boundary_uniform}).

\item  At large $|\vec{x}\,|$ the reflected field obeys the out-going radiation condition which is legitimate to choose in the form
\begin{equation}
\lim_{R\to \infty} \int_{C_R} \left |\partial_r \Psi^{\mathrm{ref} } -\mathrm{i}  k  \Psi^{\mathrm{ref}} \right |^2 \mathrm{d}s=0 
\label{radiation} 
\end{equation}
where the integration is performed over a big circle $C_R$ of radius $R$  which includes all vortices and $s$ is the length along this circle (see dashed circle in Fig.~\ref{contour}.

\item   Vortices are  considered to be impenetrable. It means that the full wave function tends to zero at  vortex positions. More precisely, in a small vicinity of each vortex the full wave function should have the following behavior
\begin{equation}
\Psi(\vec{x}\, )\underset{x\to L_j}{\longrightarrow} a_j(x-L_j+\mathrm{i}y)^{\alpha_j}+b_j(x-L_j-\mathrm{i}y)^{1-\alpha_j}\ .
\label{asymptotic_j}
\end{equation} 
 As the integer part of flux does not change boundary conditions we may and will consider fluxes in the interval $0<\alpha_j<1$.  
\end{enumerate}
From physical considerations it is clear that the solution obeying all conditions (a)-(e) is unique or, which is the same, if a function fulfills all these conditions  with  zero in-going incident field, it is identical zero. The proof of this fact can be done by a generalization of  usual  arguments developed for diffraction problems (see e.g. \cite{schot} and references therein). For completeness, we present in Appendix~\ref{uniqueness} a brief demonstration of uniqueness stressing the necessity of all requirements (a)-(e).

The usual way of solving the  problem  with two vortices is to represent the reflected outgoing field as a sum of single and double layers along the cuts
\begin{equation}
\Psi^{\mathrm{ref}}(x,y)= \int_{\mathrm{cuts}}  H_0^{(1)}\left (k\sqrt{(x-t)^2+y^2}\, \right )\mu(t)\mathrm{d}t+
\partial_y\int_{\mathrm{cuts}}  H_0^{(1)}\left (k\sqrt{(x-t)^2+y^2}\,\right )\nu(t)\mathrm{d}t\ ,
\label{general}
\end{equation}
where $ H_0^{(1)}(x)$ are the Hankel function of the first kind and zero order. 

Functions  $\mu(t)$ and $\nu(t)$ are piece-wise functions on the cuts which have to be determined from the boundary conditions  \eqref{boundary_uniform}. The   calculations are standard  (cf. e.g. \cite{optics}) and lead to the following system of equations
\begin{eqnarray}
\nu(x)&=& \tfrac{1}{2}\tan \pi \chi(x)  \int_{\mathrm{cuts}}  H_0^{(1)}(k|x-t|)\mu(t) \mathrm{d}t + \mathcal{F}(x,0)\ , \label{eq_nu}\\
\mu(x) &=& -\tfrac{1}{2}\tan \pi \chi(x) \left ( \partial_x^2 +k^2\right ) \int_{\mathrm{cuts}}   H_0^{(1)}(k|x-t|)\nu (t)\mathrm{d}t +\partial_y \mathcal{F}(x,0) \ ,  \label{eq_mu}
\end{eqnarray}
where (provided that the incident wave $\Psi^{\mathrm{inc}}(\vec{x}\,)$ has no phase jumps) 
\begin{equation}
\mathcal{F}(x,0)=  \tfrac{1}{2} \tan \pi \chi(x) \, \Psi^{\mathrm{inc}}(x,0), \qquad  \partial_y \mathcal{F}(x,0)=  \tfrac{1}{2} \tan \pi \chi(x)\,  \partial_y \Psi^{\mathrm{inc}}(x,0)\ .
\end{equation}
This approach is well suited for numerical calculations. To progress in analytic treatment,  we generalize in the next Section  the method of Ref.~\cite{myers} which has been developed for the diffraction on a finite slit.     

\section{Green function for two-vortex problem}\label{green_function}

To calculate the Green function for the two-vortex problem it is necessary to fix the incident field as in Eq.~\eqref{free_hankel}. Due to dependence of the source coordinates, $\vec{x}^{\, \prime}$, the asymptotics of the Green function denoted by  
$G(\vec{x} ,\vec{x}^{\, \prime}) $ is 
\begin{equation}
G(\vec{x} ,\vec{x}^{\, \prime}) \underset{x\to L_j}{\longrightarrow} 
a_j(\vec{x}^{\, \prime}) \, (x-L_j+ \mathrm{i}y)^{\alpha_j}+ 
b_j(\vec{x}^{\, \prime}) \, (x-L_j-\mathrm{i}y)^{1-\alpha_j}\ .
 \label{x_Lj_alpha_j}
\end{equation}
The choice of fractional power branch is dictated by the choice of the cuts.

Let us introduce auxiliary functions $A_j(\vec{x}\,)$ and $B_j(\vec{x}\,)$ independent of the incident field which obey all the conditions \eqref{helmholtz}-\eqref{asymptotic_j} except that at one vortex indicated by $j$ they have  the following asymptotic behavior
\begin{eqnarray}
& & A_j(\vec{x}\, )\underset{x\to L_j}{\longrightarrow}  \partial_{L_j}(x-L_j+\mathrm{i}y)^{\alpha_j}=-\frac{\alpha_i}{ (x-L_j+ \mathrm{i}y)^{1-\alpha_j}}\ ,\label{A_j}\\
& & B_j(\vec{x}\, )\underset{x\to L_j}{\longrightarrow}  \partial_{L_j}(x-L_j-\mathrm{i}y)^{1-\alpha_j}=-\frac{1-\alpha_i}{ (x-L_j+ \mathrm{i}y)^{\alpha_j}}\ .\label{B_j}
\end{eqnarray}
These functions are uniquely fixed by these conditions and will play an important role below. 

The  uniqueness of the solution together with point-like character of boundary conditions \eqref{x_Lj_alpha_j} permit to find different equations between functions $a_j, b_j,A_j,B_j$ \cite{myers}.  Indeed, assume that there exists an infinitesimal transformation $\hat{\delta}$ commuting with the Laplace operator such that the incident field is invariant, $\hat{\delta}\Psi^{\mathrm{inc}}(\vec{x}\, ) =0$. Then the function $\hat{\delta}\Psi(\vec{x}\, )$ is a solution with zero incident field. Nevertheless, such transformed solution is not zero because, in general, the symmetry transformation $\hat{\delta}$  changes the behavior near one or many vortices. But this change can be compensated by a suitable chosen linear combinations of  functions $A_j$ and $B_j$. It means that the function
\begin{equation}
\hat{\delta}\Psi(\vec{x}\, ) +\sum_{j}[c_j A_j(\vec{x}\, )+d_j B_j(\vec{x}\, )]
\end{equation}
obeys all conditions \eqref{helmholtz}-\eqref{asymptotic_j} but with zero internal field. By uniqueness it has to be identically zero,
\begin{equation}
\hat{\delta}\Psi(\vec{x}\, ) +\sum_{j}[c_j A_j(\vec{x}\, )+d_j B_j(\vec{x}\, )]\equiv 0\ .
\end{equation}
Combining together different transformations leads to sufficient number of equations which permits in the end to reconstruct the full solution.

When we are interested in the Green function for a system of vortices, the incident field \eqref{free_hankel}  has the following main symmetries
\begin{itemize}
\item change of vortex positions: $\hat{\delta}=\partial_{L_j}$,

\item translational invariance: $\hat{\delta}=\partial_{x}+\partial_{x^{\prime}},\qquad \hat{\delta}=\partial_{y}+\partial_{y^{\prime}}$, 
\item rotational invariance: $\hat{\delta}=L_{\theta}^{(j)}=x\partial_y-y\partial_x-L_j\partial_y $. 
\end{itemize}
In the subsequent Sections these transformations and their combinations are 
considered and equations for the Green function of two AB vortices are derived.

\subsection{Derivatives over vortex positions}

From \eqref{x_Lj_alpha_j} and definitions \eqref{A_j} and \eqref{B_j} it follows that the combination
\begin{equation}
\partial_{L_j} G(\vec{x} ,\vec{x}^{\, \prime})-a_j(\vec{x}^{\, \prime}) A_j(\vec{x}\, )-b_j(\vec{x}^{\, \prime}) B_j(\vec{x}\,)
\end{equation}
is zero at all vortices and, as was discussed above, is identically zero. 

Therefore  derivatives of the Green function  over vortex positions are  
\begin{equation}
\partial_{L_j} G(\vec{x} ,\vec{x}^{\, \prime})=a_j(\vec{x}^{\, \prime}) A_j(\vec{x}\, )+b_j(\vec{x}^{\, \prime}) B_j(\vec{x}\,)\ .
\label{partial_Lj_two_vortices}
\end{equation}
Assume for a moment that $0<\alpha<1/2$ then $a_j(\vec{x}^{\, \prime})$ and $b_j(\vec{x}^{\, \prime})$ can be calculated from the limits
\begin{equation}
a_j(\vec{x}^{\, \prime})=\lim_{x\to L_j} (x-L_j)^{-\alpha_j} G(\vec{x} ,\vec{x}^{\, \prime}),\qquad 
b_j(\vec{x}^{\, \prime})=\lim_{x\to L_j} (x-L_j)^{\alpha_j-1} \left [ G(\vec{x} ,\vec{x}^{\, \prime})-a_j(\vec{x}^{\, \prime})(x-L_1)^{\alpha_j}\right ]\ .
\end{equation}
Differentiating this limit over $L_2$ for $j=1$ and over $L_1$ for $j=2$ one concludes that
\begin{equation}
\partial_{L_2} \left (\begin{array}{c} a_1(\vec{x}^{\, \prime})\\b_1(\vec{x}^{\, \prime})\end{array}\right )= \left (\begin{array}{cc} \beta_1 & \epsilon_1 \\ \delta_1  &\zeta_1 \end{array}\right ) \left (\begin{array}{c} a_2(\vec{x}^{\, \prime})\\b_2(\vec{x}^{\, \prime})\end{array}\right )\  ,\qquad 
\partial_{L_1}\left (\begin{array}{c} a_2(\vec{x}^{\, \prime})\\b_2(\vec{x}^{\, \prime})\end{array}\right )= \left (\begin{array}{cc} \beta_2 & \epsilon_2 \\ \delta_2  &\zeta_2 \end{array}\right ) \left (\begin{array}{c} a_1(\vec{x}^{\, \prime})\\b_1(\vec{x}^{\, \prime})\end{array}\right )\ ,
\label{dif_a_b_two_vortices}
\end{equation}
where all matrix elements are independent on space coordinates.
 
Calculating the mixed derivatives  $\partial^2_{L_1\,L_2}G(\vec{x} ,\vec{x}^{\, \prime})$  one gets that the derivatives of $A_i$ and $B_i$ are expressed through the same constants as follows
\begin{equation}
\partial_{L_2}\left (\begin{array}{c} A_1(\vec{x}\,)\\B_1(\vec{x}\,)\end{array}\right )= \left (\begin{array}{cc} \beta_2 & \delta_2 \\ \epsilon_2  &\zeta_2 \end{array}\right ) \left (\begin{array}{c}A_2(\vec{x}\,)\\B_2(\vec{x}\,)\end{array}\right )\ ,\qquad 
\partial_{L_1}\left (\begin{array}{c} A_2(\vec{x}\,)\\B_2(\vec{x}\,)\end{array}\right )= \left (\begin{array}{cc} \beta_1 & \delta_1 \\ \epsilon_1  &\zeta_1 \end{array}\right ) \left (\begin{array}{c} A_1(\vec{x}\,)\\B_1(\vec{x}\,)\end{array}\right )\ .
\label{dif_A_B_two_vortices}
\end{equation}
\subsection{Translational invariance}

There exists a few simple consequences of translational invariance. First, constants like $\beta_j, \epsilon_j,\delta_j, \zeta_j$ depend only on the difference $L=L_1-L_2$. Second, functions depended only on $\vec{x}$ or $\vec{x}^{\,\prime}$ like  
$f_i=a_i, b_i, A_i, B_i$ obey the equation 
\begin{equation}
\left ( \partial_{L_1}+\partial_{L_2}+\partial_x \right )f_i(\vec{x})=0\ .
\label{f_i}
\end{equation}
This equation permits to calculate derivatives $\partial_{L_i}f_i$ in contrast to Eqs.~\eqref{dif_a_b_two_vortices} which determine derivatives $\partial_{L_j}f_i$ only with $j\neq i$.  

Consider now the change of two coordinates  simultaneously.  Comparing the behavior near two vortices \eqref{x_Lj_alpha_j} one concludes that 
\begin{equation}
\mathrm{i}\left ( \partial_y+\partial_{y^{\prime}}\right )G(\vec{x} ,\vec{x}^{\, \prime})=a_1(\vec{x}^{\, \prime}) A_1(\vec{x}\, )-b_1(\vec{x}^{\, \prime}) B_1(\vec{x}\,)+a_2(\vec{x}^{\, \prime}) A_2(\vec{x}\, )-b_2(\vec{x}^{\, \prime}) B_2(\vec{x}\,) \ .
\label{dy_dy_prime}
\end{equation}
Differentiating  this expression by $L_1$  and using  Eqs.~\eqref{dif_A_B_two_vortices} and \eqref{f_i} leads to the  identity valid for all $\vec{x}$ and $\vec{x}^{\,\prime}$
\begin{eqnarray}
& & A_1(\vec{x}\,) (\partial_{x^{\prime}}+\mathrm{i}\partial_{y^{\prime}})a_1(\vec{x}^{\, \prime})+a_1(\vec{x}^{\, \prime}) (\partial_{x}+\mathrm{i}\partial_{y})A_1(\vec{x}\,)-B_1(\vec{x}\,) (\partial_{x^{\prime}}-\mathrm{i}\partial_{y^{\prime}})b_1(\vec{x}^{\, \prime})-b_1(\vec{x}^{\, \prime}) (\partial_{x}-\mathrm{i}\partial_{y})B_1(\vec{x}\,)= \nonumber \\ 
& -&A_1(\vec{x}\,)[\beta_1 a_2(\vec{x}^{\, \prime})+\epsilon_1 b_2(\vec{x}^{\, \prime})]-a_1(\vec{x}^{\, \prime})[\beta_2 A_2(\vec{x}\,)+\delta_2 B_2(\vec{x}\,)]+A_2 (\vec{x}\,)[\beta_2 a_1(\vec{x}^{\, \prime})+\epsilon_2 b_1(\vec{x}^{\, \prime})]\nonumber\\
& + &a_2(\vec{x}^{\, \prime}) [\beta_1 A_1(\vec{x}\,)+\delta_1 b_1(\vec{x}^{\, \prime})]-B_2(\vec{x}\,) [\delta_2 a_1(\vec{x}^{\, \prime})+\zeta_2 b_1(\vec{x}^{\, \prime})]-b_2(\vec{x}^{\, \prime}) [\epsilon_1 A_1(\vec{x}\,)+\zeta_1 B_1(\vec{x}\,)] \ .
\end{eqnarray}
Substituting in this expression the most general linear relations between  derivatives and unknown  functions 
\begin{eqnarray}
& &\left ( \partial_{x}+\mathrm{i} \partial_{y}\right )A_1(\vec{x}\, )=g_1 A_1(\vec{x}\, )+ g_2 B_1(\vec{x}\, )+h_1 A_2(\vec{x}\, ) +h_2 B_2(\vec{x}\, )\ , \nonumber \\
& &\left ( \partial_{x}-\mathrm{i} \partial_{y}\right )B_1(\vec{x}\, )=g_3 A_1(\vec{x}\,  )+ g_4 B_1(\vec{x} \, )+h_3 A_2(\vec{x}\, )+h_4 B_2(\vec{x}\, )\ , \nonumber \\
& &\left ( \partial_{x^{\prime} }+\mathrm{i} \partial_{y^{\prime} }\right )a_1(\vec{x}^{\, \prime} )=G_1 a_1(\vec{x}^{\, \prime} )+G_2 b_1(\vec{x}^{\, \prime} )
+H_1  a_1(\vec{x}^{\, \prime} ) +H_2  b_2(\vec{x}^{\, \prime} ) \ ,\label{the_most_general} \\
& &\left ( \partial_{x^{\prime} }-\mathrm{i} \partial_{y^{\prime} }\right )b_1(\vec{x}^{\, \prime} )=G_3 a_1(\vec{x}^{\, \prime} )+G_4 b_1(\vec{x}^{\, \prime} )
+H_3  a_1(\vec{x}^{\, \prime} ) +H_4  b_2(\vec{x}^{\, \prime} ) \ ,\nonumber
\end{eqnarray}
and collecting identical terms proves the following formulae 
\begin{eqnarray}
& &\left ( \partial_{x}+\mathrm{i} \partial_{y}\right )A_1(\vec{x}\, )=g_1 A_1(\vec{x}\, )+ g_2 B_1(\vec{x}\, )-2\delta_2 B_2(\vec{x}\, )\ ,\nonumber  \\
& &\left ( \partial_{x}-\mathrm{i} \partial_{y}\right )B_1(\vec{x}\, )=g_3 A_1(\vec{x}\,  )+ g_4 B_1(\vec{x} \, )-2\epsilon_2 A_2(\vec{x}\, )\ ,\nonumber  \\
& &\left ( \partial_{x^{\prime}}+\mathrm{i} \partial_{y^{\prime}}\right )a_1(\vec{x}^{\, \prime})=-g_1\,  a_1(\vec{x}^{\, \prime})+
g_3\,   b_1(\vec{x}^{\, \prime})-2\epsilon_1 b_2(\vec{x}^{\, \prime})\ ,\label{translation_1}\\
& &\left ( \partial_{x^{\prime}}-\mathrm{i} \partial_{y^{\prime}}\right )b_1(\vec{x}^{\, \prime})=g_2 \,  a_1(\vec{x}^{\, \prime})
-g_4 \,  b_1(\vec{x}^{\, \prime})-2\delta_1 a_2(\vec{x}^{\, \prime})\ . \nonumber 
\end{eqnarray}
Similarly, differentiating Eq.~\eqref{dy_dy_prime} by $L_2$ gives rise to the relations
\begin{eqnarray}
& &\left ( \partial_{x}+\mathrm{i} \partial_{y}\right )A_2(\vec{x}\, )=f_1 A_2(\vec{x}\, )+f_2 B_2(\vec{x}\, )-2\delta_1 B_1(\vec{x}\, )\ ,\nonumber \\
& &\left ( \partial_{x}-\mathrm{i} \partial_{y}\right )B_2(\vec{x}\, )=f_3 A_2(\vec{x}\,  )+f_4 B_2(\vec{x}\,  )-2\epsilon_1 A_1(\vec{x}\,  )\ ,\nonumber \\
& &\left ( \partial_{x^{\prime}}+\mathrm{i} \partial_{y^{\prime}}\right )a_2(\vec{x}^{\, \prime})=-f_1\,   a_2(\vec{x}^{\, \prime})+
f_3 \,  b_2(\vec{x}^{\, \prime})-2\epsilon_2 b_1(\vec{x}^{\, \prime})\ ,\label{translation_2}\\
& &\left ( \partial_{x^{\prime}}-\mathrm{i} \partial_{y^{\prime}}\right )b_2(\vec{x}^{\, \prime})=f_2\,   a_2(\vec{x}^{\, \prime})+
f_4 \,  b_2(\vec{x}^{\, \prime})-2\delta_2 a_1(\vec{x}^{\, \prime}) \ .\nonumber 
\end{eqnarray}
Here  $g_j$ and $f_j$ with $j=1,2,3,4$ are undetermined constants depended only on $L=L_1-L_2$. 


\subsection{Rotational invariance}

The mutual rotation of $\vec{x}$ and $\vec{x}^{\, \prime}$ around any vortex is evidently the invariance transformation for the incident field \eqref{free_hankel}. Such rotations are generated by operators 
\begin{equation}
L_{\theta}^{(j)}=x\partial_y-y\partial_x-L_j\partial_y\ ,\qquad  L_{\theta^{\prime}}^{(j)}=x^{\prime} \partial_{y^{\prime}}-y^{\prime} \partial_{x^{\prime}}-L_j\partial_{y^{\prime}} \ .
\end{equation}
One has
\begin{equation}
\left (L_{\theta}^{(j)}+  L_{\theta^{\prime}}^{(j)} \right ) H_0^{(1)}\left (k\sqrt{(x-x^{\prime})^2+(y-y^{\prime})^2 } \right)=0\ .
\end{equation}
Exactly as it has been done above one concludes that
\begin{eqnarray}
& &\mathrm{i}\left (L_{\theta}^{(2)}+  L_{\theta^{\prime}}^{(2)} \right )G(\vec{x} ,\vec{x}^{\, \prime}) =L\,  a_1(\vec{x}^{\, \prime})A_1(\vec{x}\, )-L\, b_1(\vec{x}^{\, \prime})B_1(\vec{x}\, ) \ ,\label{rotation_G}\\
 & &\mathrm{i}\left (L_{\theta}^{(1)}+  L_{\theta^{\prime}}^{(1)} \right )G(\vec{x} ,\vec{x}^{\, \prime}) =L\,  a_2(\vec{x}^{\, \prime})A_2(\vec{x}\, )-L\, b_2(\vec{x}^{\, \prime})B_2(\vec{x}\, ) \nonumber
\end{eqnarray}
with $L=L_1-L_2$.
 
Differentiating these equations by $L_j$ one finds the following  relations  (similar equations for functions $a_i$ and $b_j$  are not presented) 
\begin{equation}
\mathrm{i}L_{\theta}^{(1)} \left ( \begin{array}{c}A_1\\B_1 \end{array} \right )=M_1 \left ( \begin{array}{c}A_1\\B_1 \end{array} \right ) +L \left (\begin{array}{cc} -\beta_2 & \delta_2\\-\epsilon_2 & \zeta_2\end{array}  \right )\left ( \begin{array}{c}A_2\\ B_2 \end{array}\right ),\qquad 
\mathrm{i}L_{\theta}^{(2)} \left ( \begin{array}{c}A_2\\B_2 \end{array} \right )=M_2 \left ( \begin{array}{c}A_2\\B_2 \end{array} \right ) +L \left (\begin{array}{cc} \beta_1 & -\delta_1\\ \epsilon_1 & -\zeta_1\end{array}  \right )\left ( \begin{array}{c}A_1\\B_1 \end{array}\right )\ , \label{rotation_2} 
\end{equation}
where $M_j$ are $2\times 2$ matrices with undetermined coefficients. 

\section{Meaning and determination of coefficients}\label{determination_constants}

Eqs.~\eqref{translation_1}, \eqref{translation_2}  and  \eqref{rotation_2} are concise consequences of symmetries of the incident field \eqref{free_hankel}. But these equations contain many coefficients depended, in general, on the distance between vortices, $L$.  

The first simplification comes from the fact that 
functions $a_j$ and $b_j$ for two vortices can be expressed through functions $A_j$ and $B_j$ by  the reciprocity relation \eqref{full_reciprocity} proven in Appendix~\ref{uniqueness}. It states
\begin{equation}
G(\vec{x},\vec{x}^{\, \prime})=G(\hat{S}\vec{x}^{\, \prime},\hat{S}\vec{x}\,),\qquad \hat{S}(x,y)=(x,-y)\ . 
\end{equation}  
Using Eqs.~\eqref{partial_Lj_two_vortices} and \eqref{rotation_G} and taking into account that  
$ L_{\theta}^{(j)} F(\hat{S}\vec{x})=-\hat{S}L_{\theta}^{(j)}  F(\vec{x})$ one concludes that 
\begin{equation}
\left (\begin{array}{c} A_1(\vec{x}\, )\\B_1(\vec{x}\,)\end{array}\right )= \left (\begin{array}{cc} 0& t_1 \\ t_1 &0 \end{array}\right )\left (\begin{array}{c} a_1(S\vec{x}\, )\\b_1(S\vec{x}\,)\end{array}\right ),\qquad
 \left (\begin{array}{c} A_2(\vec{x}\, )\\B_2(\vec{x}\,)\end{array}\right )=\left (\begin{array}{cc} 0 & t_2 \\ t_2 &0 \end{array}\right )\left (\begin{array}{c} a_2(S\vec{x}\, )\\b_2(S\vec{x}\,)\end{array}\right )\ .
\label{A_a} 
\end{equation}
Substituting these values to Eqs.~\eqref{dif_A_B_two_vortices} and comparing with Eqs.~\eqref{dif_a_b_two_vortices} gives
\begin{equation}
\left (\begin{array}{cc} \beta_2 & \epsilon_2 \\ \delta_2  &\zeta_2 \end{array}\right ) =\rho \left (\begin{array}{cc} \zeta_1 & \epsilon_1 \\ \delta_1  & \beta_1  \end{array}\right ),\qquad \rho=\frac{t_1}{t_2} \ .
\label{relation_2_1}
\end{equation}
These  relations  mean that one can consider equations only for functions $A_j$ and $B_j$. Values of constants $t_j$ and $\rho$ are calculated in Appendix~\eqref{one_vortex}.   

In the Myers method \cite{myers} used in the previous Sections constants  appeared in different formulae are just the most general coefficients of expansion (cf. e.g. Eqs.~\eqref{the_most_general}). They also can be interpreted as sub-leading terms of expansion of auxiliary functions  $A_j(\vec{x}\,)$ and $B_j(\vec{x}\, )$ in small vicinity of vortex positions
\begin{equation}
\left ( \begin{array}{c} A_1(\vec{x}\,)\\A_2(\vec{x}\,)\\B_1(\vec{x}\,)\\B_2(\vec{x}\,)\end{array}\right )\underset{x\to L_1}{\longrightarrow} \left ( \begin{array}{c} 
\partial_{L_1}(x-L_1+\mathrm{i}y)^{\alpha_1}-\tfrac{1}{2} g_2 (x-L_1-\mathrm{i}y)^{1-\alpha_1}\\
\beta_1(x-L_1+\mathrm{i}y)^{\alpha_1}+\delta_1(x-L_1-\mathrm{i}y)^{1-\alpha_1}\\
\partial_{L_1}(x-L_1-\mathrm{i}y)^{1-\alpha_1}-\tfrac{1}{2}g_3(x-L_1+\mathrm{i}y)^{\alpha_1} \\
  \epsilon_1(x-L_1+\mathrm{i}y)^{\alpha_1}+\zeta_1(x-L_1-\mathrm{i}y)^{1-\alpha_1}
\end{array}\right )\ ,
\label{expansions_AB_1}
\end{equation}
and 
\begin{equation}
\left ( \begin{array}{c} A_1(\vec{x}\,)\\A_2(\vec{x}\,)\\B_1(\vec{x}\,)\\B_2(\vec{x}\,)\end{array}\right )\underset{x\to L_2}{\longrightarrow} \left ( \begin{array}{c} 
 \beta_2(x-L_2+\mathrm{i}y)^{\alpha_2}+\delta_2(x-L_2-\mathrm{i}y)^{1-\alpha_2}\\
\partial_{L_2}(x-L_2+\mathrm{i}y)^{\alpha_2}-\tfrac{1}{2}f_2 (x-L_2-\mathrm{i}y)^{1-\alpha_2}\\
 \epsilon_1(x-L_2+\mathrm{i}y)^{\alpha_2}+\zeta_2(x-L_2-\mathrm{i}y)^{1-\alpha_2}\\
 \partial_{L_2}(x-L_2-\mathrm{i}y)^{1-\alpha_1}-\tfrac{1}{2} f_3(x-L_2+\mathrm{i}y)^{\alpha_2}   
\end{array}\right ) \ .
\label{expansions_AB_2}
\end{equation}
Postulating these relations and comparing the dominant singularities at the vortices one can prove
all Eqs.~\eqref{dif_a_b_two_vortices}, \eqref{dif_A_B_two_vortices},   \eqref{translation_1}, \eqref{translation_2},     \eqref{rotation_2}  as it has been done in \cite{swj}. 

\vspace{.4cm}
    
The first series of relations between these coefficients  is obtained by  differentiating the both sides of Eqs.~\eqref{translation_1} and \eqref{translation_2}  by $L_j$. As all derivatives are known by previous formulae one gets many interrelations between the constants.  In particular, it  follows that  
\begin{equation}
\partial_L g_1=\partial_L g_4=\partial_L f_1=\partial_L f_4=0\ .
\end{equation}
It means that constants $g_1,g_4,f_1,f_4$ are independent on $L$.  From the asymptotic behavior of two vortices at large separation given in Appendix~\ref{one_vortex} one gets that  $g_1=g_4=f_1=f_4=0$. We rewrite the above equations  with these values
\begin{equation}
( \partial_{x}+\mathrm{i} \partial_{y})\left ( \begin{array}{c} a_1(\vec{x}^{\,\prime}  ) \\ a_2(\vec{x}^{\,\prime} ) \end{array}\right )=
\left (\begin{array}{cc } g_3&-2\epsilon_1\\ -2\epsilon_2& f_3 \end{array}\right )  \left (\begin{array}{c} b_1(\vec{x}^{\,\prime} ) \\ b_2(\vec{x}^{\,\prime} )\end{array}\right )\ ,\qquad
( \partial_{x}-\mathrm{i} \partial_{y} )\left (\begin{array}{c} b_1(\vec{x}^{\,\prime} ) \\ b_2(\vec{x}^{\,\prime} ) \end{array}\right )=
 \left (\begin{array}{cc } g_2&-2\delta_1\\ -2\delta_2& f_2 \end{array}\right ) \left (\begin{array}{c} a_1(\vec{x}^{\,\prime} ) \\ a_2(\vec{x}^{\,\prime} )\end{array}\right )\  , 
 \label{translation_final_ab}
\end{equation}
and
\begin{equation}
( \partial_{x}+\mathrm{i} \partial_{y})\left ( \begin{array}{c} A_1(\vec{x}\, ) \\ A_2(\vec{x}\, ) \end{array}\right )=
\left (\begin{array}{cc } g_2&-2\delta_2\\ -2\delta_1& f_2 \end{array}\right )  \left (\begin{array}{c} B_1(\vec{x}\, ) \\ B_2(\vec{x}\, )\end{array}\right ) ,\qquad
( \partial_{x}-\mathrm{i} \partial_{y} )\left (\begin{array}{c} B_1(\vec{x}\, ) \\ B_2(\vec{x}\, ) \end{array}\right )=
 \left (\begin{array}{cc } g_3&-2\epsilon_2\\ -2\epsilon_1& f_3 \end{array}\right ) \left (\begin{array}{c} A_1(\vec{x}\, ) \\ A_2(\vec{x}\, )\end{array}\right )\ . 
 \label{translation_final}
\end{equation}
Applying $\left ( \partial_{x}\pm \mathrm{i} \partial_{y}\right )$ operator to these equations  and using the fact that each function $f(\vec{x}\,)$ has to obey the Helmholtz equation, $(\partial_x^2+\partial_y^2+k^2)f(\vec{x}\,)=0$, one finds that
\begin{equation}
g_2 g_3+4\epsilon_2 \delta_1 + k^2=0,\qquad 
\epsilon_2\, f_2+\delta_2\, g_3=0,\qquad \delta_2\,  f_3+\epsilon_2\, g_2=0,\qquad \epsilon_1 \delta_2 = \delta_1 \epsilon_2 .
\label{k_2}
\end{equation} 
Differentiating \eqref{translation_final} by $L_j$ gives
\begin{eqnarray}
& &\dot{g}_2= 2(\zeta_1\delta_2+\beta_2\delta_1),\qquad \dot{g}_3=2(\epsilon_2\beta_1+\epsilon_1\zeta_2),\qquad
 \dot{f}_2=-2(\delta_1\zeta_2+\delta_2\beta_1),\qquad \dot{f}_3=-2(\zeta_1\epsilon_2+\beta_2\epsilon_1)\nonumber\\ 
& &2\dot{\delta}_2= f_2\beta_2-g_2\zeta_2,\qquad  2\dot{\epsilon}_2=f_3\zeta_2-g_3\beta_2,\qquad 2\dot{\delta}_1=f_2\zeta_1-g_2\beta_1,\qquad 
2\dot{\epsilon}_1=f_3\beta_1-g_3\zeta_1 .
\end{eqnarray}
Here and below the dot indicates the derivative over $L$.

In a similar way the differentiation  by $L_j$ the rotation equations  \eqref{rotation_2} gets  many other relations. In particular, the both matrices $M_1$ and $M_2$ are constant. From the solution of well-separated vortices (see Appendix~\ref{one_vortex}) it follows that the both matrices are diagonal. For simplicity we impose this condition from now. Therefore the rotations have the form
\begin{eqnarray}
\mathrm{i}L_{\theta}^{(1)} A_1&=& m_1 A_1  -L \beta_2 A_2+ L\delta_2 B_2\ , \qquad 
\mathrm{i}L_{\theta}^{(1)} B_1= n_1 B_1  -L \epsilon_2 A_2+L  \zeta_2 B_2\ ,\nonumber  \\
\mathrm{i}L_{\theta}^{(2)} A_2&=& m_2 A_2 + L \beta_1 A_1-L \delta_1 B_1\ ,\label{rotation_final}\qquad 
\mathrm{i}L_{\theta}^{(2)} B_2= n_2 B_2 + L \epsilon_1 A_1- L  \zeta_1 B_1\ .
\end{eqnarray}
In the indicated way one gets the following list of equations
\begin{eqnarray}
(L\epsilon_1)^{\cdot}&=&(m_1-n_2)\epsilon_1-L\zeta_1 g_3, \qquad (L\epsilon_2)^{\cdot}= (m_2-n_1)\epsilon_2+ L\zeta_2 f_3,\nonumber\\
(L\zeta_1)^{\cdot}&=&(n_2-n_1)\zeta_1-L\epsilon_1 g_2 ,\qquad (L\zeta_2)^{\cdot}= (n_1-n_2)\zeta_2+L\epsilon_2 f_2,\nonumber\\
(L\beta_1)^{\cdot}&=&(m_1-m_2)\beta_1- L\delta_1 g_3 ,\qquad (L\beta_2)^{\cdot}=(m_2-m_1)\beta_2+ L\delta_2 f_3, \label{diff_eqs}\\
(L\delta_1)^{\cdot}&=&(m_2-n_1)\delta_1 - L\beta_1 g_2,\qquad (L\delta_2)^{\cdot}=(m_1-n_2)\delta_2+L\beta_2 f_2 .\nonumber
\end{eqnarray} 
According to Eqs.~\eqref{m_n} of Appendix~\ref{one_vortex} the values of $m_i$ and $n_i$ are 
\begin{equation}
m_1=1-\alpha_1,\qquad n_1=-\alpha_1, \qquad m_2=1-\alpha_2,\qquad n_2=-\alpha_2.
\end{equation}
As was discussed above  one has to take into account only equations related with one vortex (say $L_1$). We rewrite them again 
\begin{equation}
\dot{\epsilon_1}=\frac{\gamma}{L}\epsilon_1-\zeta_1 g_3, \qquad 
\dot{\zeta_1}= -\frac{\gamma+1}{L}\zeta_1- \epsilon_1 g_2, \qquad 
\dot{\beta_1}=\frac{\gamma-1}{L}\beta_1- \delta_1 g_3, \qquad 
\dot{\delta_1}=-\frac{\gamma}{L}\delta_1 - \beta_1 g_2 
\label{eqs_different_fluxes}
\end{equation} 
with $\gamma=\alpha_2-\alpha_1$.

To these equations one should add the following ones
\begin{equation}
g_2 g_3+4 \epsilon_2 \delta_1 + k^2=0,\qquad \beta_1 \epsilon_2 g_2-\beta_2\delta_1 g_3 +\frac{2\gamma}{L}\delta_2 \epsilon_1=0.
\label{constants}
\end{equation} 
The second of these equations is a consequence of Eqs.~\eqref{diff_eqs} when relations \eqref{relation_2_1} are imposed. 

It is convenient to introduce the following notations 
\begin{equation}
y=\delta_2 \epsilon_1 =\delta_1 \epsilon_2 ,\qquad v=f_3 \delta_2 =-g_2\epsilon_2 ,\qquad w=g_3\delta_1 =-f_2\epsilon_1.
\label{variables}
\end{equation}
Direct check proves that these variables obey the equations 
\begin{equation}
\dot{y}= \beta_1 v-\beta_2 w,\quad 
\dot{\beta}_1=\frac{\gamma-1}{L}\beta_1 -w,\quad
\dot{\beta}_2=-\frac{\gamma+1}{L}\beta_2 +v, \quad
\dot{w}=-\frac{\gamma}{L}w+\beta_1(k^2+8 y),\quad
\dot{v}=\frac{\gamma}{L}v-\beta_2(k^2+8 y)
\label{y_beta_u_w}
\end{equation}
with $\gamma=m_1-m_2=\alpha_2-\alpha_1$. 

Eqs.~\eqref{constants} are equivalent to the existence of  two integrals of motion 
\begin{equation}
v w-y(4y+k^2)=0, \qquad \beta_1 v+\beta_2 w -\frac{2\gamma}{L} y=0 .
\label{integrals}
\end{equation}
Introducing new variable, $z=\beta_1\beta_2$, one gets the following equations
\begin{equation}
\dot{z}=-\frac{2}{L}z+\dot{y},\qquad \beta_1 v=\frac{\gamma}{L}y+\frac{1}{2}\dot{y},\qquad \beta_2 w=\frac{\gamma}{L}y -\frac{1}{2}\dot{y}.
\end{equation}
Combining with Eqs.~\eqref{integrals} leads to the final system of equations
\begin{equation}
\dot{z}=-\frac{2}{L}z+\dot{y},\qquad \frac{1}{4}\dot{y}^2-\frac{\gamma^2}{L^2}y^2+zy(4y+k^2)=0.
\label{equations_z_y}
\end{equation} 
Substitutions
\begin{equation}
 y=-\frac{k^2}{4}Y,\qquad z=-\frac{Z-\gamma^2}{4 L^2  },\qquad L=\frac{x}{k} 
 \label{rescaling}
\end{equation} 
transform Eqs.~\eqref{equations_z_y} to the form
\begin{equation}
\frac{x^2}{4}Y^{\prime\, 2}-\gamma^2 Y= Z Y (Y-1),\qquad Z^{\prime}=x^2 Y^{\prime}.
\end{equation}
Removing $Z$ gives one non-linear equation for $Y$
\begin{equation} 
Y^{\prime\prime} = \left ( \dfrac{1}{2Y}+\dfrac{1}{2(Y-1)}\right )Y^{\prime \, 2}-\dfrac{Y^{\prime}}{x}+2Y(Y-1)-
\dfrac{2\gamma^2 Y}{x^2(Y-1)}\ .
\label{final_equation}
\end{equation}
Finally one more change of variables 
\begin{equation}
Y=\frac{V}{V-1},\qquad x=\sqrt{t}
\end{equation}
transforms this equation to the canonical form of Painlev\'e V equation   (see e.g. \cite{gambier})
\begin{equation}
\frac{ \mathrm{d}^2 V}{\mathrm{d} t^2} = 
\Big ( \frac{1}{2V}+\frac{1}{V-1} \Big ) \Big (\frac{\mathrm{d} V}{\mathrm{d} t}\Big )^2 
-\frac{1}{t} \frac{\mathrm{d} V}{\mathrm{d} t} +\frac{\gamma^2 V(V-1)^2 }{2t^2}\ .
\label{main_equation}
\end{equation}
General Painlev\'e V equation \cite{gambier} is
\begin{equation}
y^{\prime \prime}=y^{\prime 2}\left ( \frac{1}{2y}+\frac{1}{y-1}\right )-\frac{y^{\prime}}{t}+\frac{(y-1)^2}{t^2}\left [ \alpha y +\frac{\beta}{y}\right ]+\epsilon \frac{y}{t}+\frac{\delta y(y+1)}{y-1}\ .
\end{equation}
Therefore Eq.~\eqref{main_equation}  is the Painlev\'e V equation with the parameters
\begin{equation}
\alpha=\tfrac{1}{2}\gamma^2,\qquad \epsilon=-\tfrac{1}{2},\qquad \beta=0,\qquad \delta=0.
\end{equation}
The Painlev\'e V equation with $\delta=0$ can be reduced to the Painlev\'e III equation \cite{gromak}. Consider the following B\"acklund-type transformation for two functions $Y$ and $W$
\begin{equation}
\dot{Y }= \frac{2\gamma }{x}Y-\frac{2Y(Y-1)}{W}, \qquad
\dot{W}= 1-2Y+\frac{1+2\gamma}{x}W+W^2\ .
\end{equation}
Finding $W$ from the first equation and substituting it into the second one leads to Eq.~\eqref{final_equation}  for $Y$. 
Calculating $Y$ from the second equation and putting it to the first one gives 
\begin{equation}
\ddot{W}=\frac{\dot{W}^2}{W}-\frac{\dot{W}}{x}+W^3 +\frac{2(1+\gamma)W^2-2 \gamma}{x}-\frac{1}{W}\   .
\label{w_equation}
\end{equation}
Standard form of the Painlev\'e III equation is (see e.g. \cite{gambier})
\begin{equation}
\ddot{y}=\frac{\dot{y}^2}{y}-\frac{\dot{y}}{x}+\alpha y^3 +\frac{\beta y^2 +\epsilon}{x}+\frac{\delta}{y} \ .
\end{equation}
Therefore Eq.~\eqref{w_equation}  is the Painlev\'e III equation with parameters
\begin{equation}
\alpha=1,\qquad \beta =2(1+\gamma),\qquad \epsilon =-2\gamma,\qquad \delta=-1 .
\end{equation}
The knowledge of the solution $y=y(L)$  permits to find all other quantities by simple integration. First, $z(L)$ is determined directly from the second of Eqs.~\eqref{equations_z_y}. From  Eqs.~\eqref{y_beta_u_w}  it follows that 
\begin{eqnarray}
u&=& \sqrt{y(4y+k^2)}\, \mathrm{e}^{-S}, \qquad w=\sqrt{y(4y+k^2)}\, \mathrm{e}^{S},\\
\beta_1&=&\dfrac{\gamma y/L+\dot{y}/2}{\sqrt{y(4y+k^2)}}\, \mathrm{e}^{S}, \
\beta_2=\dfrac{\gamma y/L-\dot{y}/2}{\sqrt{y(4y+k^2)}}\, \mathrm{e}^{-S} 
\end{eqnarray}
where function $S=S(L)$ is calculated from  the equation
\begin{equation}
\dot{S}=\frac{4\gamma y}{L(k^2+4y)}\ .
\end{equation}
Using Eqs.~\eqref{eqs_different_fluxes} one demonstrates that 
\begin{equation}
\epsilon_1=C\sqrt{y},\qquad \delta_1=\frac{1}{\rho C}\sqrt{y},\qquad \epsilon_2=\rho C\sqrt{y},\qquad \delta_2=\frac{1}{ C}\sqrt{y}
\end{equation}
where $C=C(\alpha_1,\alpha_2)$ is a constant which can be calculated from limiting values found in Section~\ref{small_distance}
\begin{equation}
C(\alpha_1,\alpha_2)=\left(\frac{k}{2} \right )^{\alpha_1+\alpha_2-1}\frac{\Gamma(2-\alpha_2)}{\Gamma(1+\alpha_1)}\sqrt{\frac{\sin \pi \alpha_2}{\sin \pi\alpha_1}}\mathrm{e}^{-\pi\mathrm{i}(3\alpha_1+\alpha_2)/2}\ .
\end{equation}
Other quantities can be calculated from the definition and reciprocity relation \eqref{relation_2_1} 
\begin{equation}
\zeta_1=\frac{1}{\rho}\beta_2,\qquad \zeta_2=\rho \beta_1,\qquad g_2=-\frac{v}{\epsilon_2},\qquad g_3=\frac{w}{\delta_1}.
\end{equation}


\section{Scattering amplitude}\label{scattering_amplitude}

A typical scattering problem consists in the determination of wave function, $\Psi^{\mathrm{scat}}(\vec{x}\,)$,  when the incident field is chosen as plane wave \eqref{plane_wave}.  The main physical quantity of interest is the scattering amplitude, $\mathcal{F}(\theta,\phi)$, obtained  from the asymptotic behavior of this function
\begin{equation}
\Psi^{\mathrm{scat}}(\vec{x}\,)\underset{|\vec{x}\,|\to\infty }{\longrightarrow} \mathrm{e}^{\mathrm{i}k r \cos (\theta-\phi)}+\sqrt{\frac{2}{\pi \mathrm{i} k r}} \mathrm{e}^{\mathrm{i}kr} \mathcal{F}(\theta,\phi).
\label{scattering_function}
\end{equation}
Here $\vec{x}=(r\cos \theta, r\sin \theta)$ and angle $\phi$ determines the direction of the incident plane wave. 

Let us consider the limit of the Green function, $G(\vec{x},\vec{x}^{\,\prime})$, with $\vec{x}^{\,\prime}=(R\cos \phi, R\sin \phi)$ and $R\to \infty$. As  
\begin{equation}
\lim_{R\to\infty} \frac{ H_0^{(1)}(k|\vec{x}-\vec{x}^{\, \prime}|)}{H_0^{(1)}(k|\vec{x}^{\, \prime}|)} =\mathrm{e}^{-\mathrm{i}kr\cos(\theta-\phi)},
\end{equation}
the scattering wave function with asymptotics \eqref{scattering_function} can be extracted from the Green function as follows  (cf. \cite{myers})
\begin{equation}
\Psi^{\mathrm{scat}}(\vec{x}\,)=\lim_{R\to\infty} \frac{G(\vec{x},-\vec{x}^{\,\prime})}{\tfrac{1}{4\mathrm{i}}H_0^{(1)}(k|\vec{x}^{\, \prime}|) }\ .
\label{lim_G}
\end{equation}
Functions $A_j(\vec{x}\,)$ and $B_j(\vec{x}\,)$ defined in the previous Section (cf. Eqs.~\eqref{A_j} and \eqref{B_j}) obey the Helmholtz equation and due to the radiation condition \eqref{radiation} have the  following asymptotic behavior
\begin{equation}
A_j(\vec{x}\,)\underset{|x|\to\infty }{\longrightarrow} \sqrt{\frac{2}{\pi \mathrm{i} k r}} \mathrm{e}^{\mathrm{i}kr} F_j(\theta)\mathrm{e}^{-\mathrm{i} k\cos \theta L_j } ,\qquad B_j(\vec{x}\,)\underset{|x|\to\infty }{\longrightarrow} \sqrt{\frac{2}{\pi \mathrm{i}k r}} \mathrm{e}^{\mathrm{i}kr} G_j(\theta)\mathrm{e}^{-\mathrm{i} k\cos \theta L_j }
\label{F_j_G_j}
\end{equation} 
with certain functions $F_j(\theta) $ and $G_j(\theta) $.

Because of translational invariance $(\partial_{L_1}+\partial_{L_2}+\partial_{x}) A_j=0$, $ (\partial_{L_1}+\partial_{L_2}+\partial_{x}) B_j=0$, functions $F_j(\theta)\equiv F_j(\theta,L)$ and $G_j(\theta)\equiv G_j(\theta, L)$  depend only on the distance between vortices, $L=L_1-L_2$ (and angle $\theta$). They have the meaning of asymptotics \eqref{F_j_G_j} when center of polar coordinates is chosen at  vortex $j$. To simplify notations,  the arguments of $F_j$ and $G_j$ are dropped when it will not lead to confusion. 

For the scattering on AB vortices the exact wave function $\Psi(\vec{x}\,)$  tends to zero at the vortex positions (cf. \eqref{asymptotic_j})
\begin{equation}
\Psi^{\mathrm{scat}}(\vec{x}\, )\underset{x\to L_j}{\longrightarrow} \tilde{a}_j(x-L_j+\mathrm{i}y)^{\alpha_j}+\tilde{b}_j(x-L_j-\mathrm{i}y)^{1-\alpha_j}\ .
\end{equation}
It is plain that the operator $\hat{\delta}=\partial_x-\mathrm{i}k\cos \phi$ gives zero when acting on the incident plane wave in Eq.~\eqref{scattering_function}. Therefore $\hat{\delta}\Psi^{\mathrm{scat}}(\vec{x}\, )$ corresponds to the zero incident field. Comparing the behavior near the vortices one gets that 
\begin{equation}
(\partial_x-\mathrm{i}k\cos \phi)\Psi^{\mathrm{scat}}(\vec{x}\, )+\sum_j\left [ \tilde{a}_j A_j(\vec{x}\,)+\tilde{b}_j B_j(\vec{x}\,)\right ]=0 .
\label{op_scat}
\end{equation} 
Applying this relation to  asymptotic expression \eqref{scattering_function} one concludes that
\begin{equation}
\mathrm{i}k(\cos \theta -\cos \phi)\mathcal{F}(\theta,\phi)=-\sum_{j} \left [ \tilde{a}_j F_j(\theta)+\tilde{b}_j G_j(\theta)\right ]\mathrm{e}^{-\mathrm{i} k\cos \theta L_j }\ .
\label{embedding}
\end{equation}
Such type of expressions is called embedding formulae in the theory of diffraction \cite{shanin}.

Values of $\tilde{a}_j$ and $\tilde{b}_j$ can be calculated from the limit \eqref{lim_G} together with \eqref{op_scat} and the reciprocity conditions \eqref{A_a} 
\begin{equation}
\tilde{a}_j=\lim_{R\to\infty} \frac{a_j(-\vec{x}^{\,\prime})}{\tfrac{1}{4\mathrm{i}}H_0^{(1)}(k|\vec{x}^{\, \prime}|) }=\frac{4\mathrm{i}}{t_j}G_j(\pi-\phi)\mathrm{e}^{\mathrm{i} k\cos \phi L_j },\qquad \tilde{b}_j=\lim_{R\to\infty} \frac{b_j(-\vec{x}^{\,\prime})}{\tfrac{1}{4\mathrm{i}}H_0^{(1)}(k|\vec{x}^{\, \prime}|) }=\frac{4\mathrm{i}}{t_j}F_j(\pi-\phi)\mathrm{e}^{\mathrm{i} k\cos \phi L_j }.
\end{equation}
For clarity the argument of $G_j$ and $F_j$ functions here is written as $\pi -\phi$. In general the choice of the branch has to be consistent with the position of the cut. 
  
Finally one obtains that the AB scattering amplitude is 
\begin{equation}
 \mathcal{F}(\theta,\phi)=-\frac{4}{k(\cos \theta -\cos \phi)}\sum_{j} \frac{1}{t_j}\Big[ G_j(\pi-\phi) F_j(\theta)+F_j(\pi-\phi) G_j(\theta)\Big ]\mathrm{e}^{\mathrm{i} k L_j ( \cos \phi-\cos \theta)}.
\label{f_g} 
\end{equation}
The relations derived in previous Sections induce  equations for $F_j$ and $G_j$. From Eqs.~\eqref{translation_final} and it follows that  functions $G_j$ are linear combinations of $F_j$ 
\begin{equation}
\left ( \begin{array}{c}G_1\\G_2 \end{array}\right )=V_1 \left ( \begin{array}{c}F_1\\F_2 \end{array}\right )\ ,\qquad \left ( \begin{array}{c}F_1\\F_2 \end{array}\right )=V_2 \left ( \begin{array}{c}G_1 \\ G_2 \end{array}\right )\, 
\label{g_f}
\end{equation}
where 
\begin{equation}
 V_1=\frac{\mathrm{e}^{\mathrm{i}\theta}}{\mathrm{i}k}
\left ( \begin{array}{c c}g_3& -2\epsilon_2\, \mathrm{e}^{\mathrm{i} kL\cos \theta} \\-2\epsilon_1\, \mathrm{e}^{-\mathrm{i} kL\cos \theta} & f_3 \end{array}\right )\ ,\qquad 
V_2=\frac{\mathrm{e}^{-\mathrm{i}\theta}}{\mathrm{i}k}
\left ( \begin{array}{c c}g_2& -2\delta_2\, \mathrm{e}^{\mathrm{i} kL\cos \theta} \\ -2\epsilon_1\, \mathrm{e}^{-\mathrm{i} kL\cos \theta} & f_2\end{array}\right )\ .
\end{equation}
Conditions \eqref{k_2} imply that  $V_1 V_2=1$.

Eqs.~\eqref{dif_A_B_two_vortices} and \eqref{rotation_final} signify that derivatives of $F_j$ and $G_j$ over $L$ and $\theta$  obey the equations
\begin{equation}
\partial_{L} \left (\begin{array}{c} F_1\\G_1\end{array}\right )= -\mathrm{e}^{\mathrm{i} kL\cos \theta}\left (\begin{array}{cc} \beta_2 & \delta_2 \\ \epsilon_2  &\zeta_2 \end{array}\right ) \left (\begin{array}{c}F_2\\G_2\end{array}\right ),\qquad 
\partial_{L} \left (\begin{array}{c} F_2\\G_2\end{array}\right )=\mathrm{e}^{-\mathrm{i} kL\cos \theta} \left (\begin{array}{cc} \beta_1 & \delta_1 \\ \epsilon_1  &\zeta_1 \end{array}\right ) \left (\begin{array}{c} F_1\\G_1\end{array}\right ),
\label{diff_L}
\end{equation}
and 
\begin{eqnarray}
& & \mathrm{i}\partial_{\theta} \left (\begin{array}{c} F_1\\G_1\end{array}\right )
=\left (\begin{array}{cc} m_1& 0 \\ 0  & n_1 \end{array}\right ) \left (\begin{array}{c}F_1\\G_1\end{array}\right )+ 
L\mathrm{e}^{\mathrm{i} kL\cos \theta}  \left (\begin{array}{cc}  -\beta_2 & \delta_2   \\  -\epsilon_2 &\zeta_2 \end{array}\right ) \left (\begin{array}{c}F_2\\G_2\end{array}\right ),\\ 
& &\mathrm{i}\partial_{\theta} \left (\begin{array}{c} F_2\\G_2\end{array}\right )
=\left (\begin{array}{cc} m_2& 0 \\ 0  & n_2 \end{array}\right ) \left (\begin{array}{c}F_2\\G_2\end{array}\right )+ 
L \mathrm{e}^{-\mathrm{i} kL\cos \theta}\left (\begin{array}{cc}  \beta_1 & -\delta_1   \\  \epsilon_1 &-\zeta_1 \end{array}\right ) \left (\begin{array}{c}F_1\\G_1\end{array}\right ).
\end{eqnarray}
Expressing $G_j$ through $F_j$ by Eqs.~\eqref{g_f} one finds 
\begin{equation}
\partial_L \left ( \begin{array}{c}F_1 \\F_2  \end{array}\right )=M \left ( \begin{array}{c}F_1\\F_2 \end{array}\right ),\qquad \mathrm{i}\partial_{\theta}\left ( \begin{array}{c}F_1\\F_2 \end{array}\right )=N \left ( \begin{array}{c}F_1\\F_2 \end{array}\right )\ ,
\end{equation}
where matrices $M$ and $N$ are 
\begin{eqnarray}
 M&=& \left ( \begin{array}{l r}  -2\mathrm{i} y\dfrac{ \mathrm{e}^{\mathrm{i}\theta}}{k} & 
 -\left[\beta_2  - \mathrm{i} v \dfrac{\mathrm{e}^{\mathrm{i}\theta}}{k}\right ] \mathrm{e}^{\mathrm{i} kL\cos \theta}  \\
\left[ \beta_1  - \mathrm{i} w \dfrac{ \mathrm{e}^{\mathrm{i}\theta}}{k}\right ] \mathrm{e}^{-\mathrm{i} kL\cos \theta}   & 
2\mathrm{i} y \dfrac{\mathrm{e}^{\mathrm{i}\theta}}{k}
\end{array}\right ),\\
N&=& \left ( \begin{array}{l r} 
m_1 +2\mathrm{i} y \dfrac{\mathrm{e}^{\mathrm{i}\theta}}{k} & 
 - L \left [  \beta_2  +\mathrm{i} v \dfrac{\mathrm{e}^{\mathrm{i}\theta}}{k}\right ] \mathrm{e}^{\mathrm{i} kL\cos \theta}  \\
L \left [ \beta_1  + \mathrm{i} w \dfrac{\mathrm{e}^{\mathrm{i}\theta}}{k}\right ] \mathrm{e}^{-\mathrm{i} kL\cos \theta}  & 
m_2 -2 \mathrm{i} y \dfrac{\mathrm{e}^{\mathrm{i}\theta}}{k}
\end{array}\right ).
\end{eqnarray}
Here the  notations are  the same as in Eqs.~\eqref{variables}. 

The compatibility condition
\begin{equation}
\partial_L N-\mathrm{i}\partial_{\theta} M=MN-NM
\end{equation}
is equivalent to Eqs.~\eqref{y_beta_u_w} which is another way of their derivation.  


\section{Solution  at small and large vortex separation}\label{small_distance}

To really use equations derived in the previous Sections it is necessary to know the values of all variables at a certain point. In this Section it is demonstrated how to  find wave function and scattering amplitude  for two AB vortices  when the separation between them, $L$, is small and large  with respect to the wavelength. The cases of small-vortex separation  with opposite fluxes (i.e. $\alpha_2=1-\alpha_1$) and with two arbitrary fluxes $\alpha_1$ and $\alpha_2$ require different arguments  and are discussed separately. 


\subsection{Two vortices with opposite  fluxes at small distances}

The vortices with opposite fluxes is considered first. Let the vortex with flux $\alpha$  be in the point $L_1=L$  and the second vortex with opposite flux be at $L_2=0$  ($0<\alpha<1$).   

The full wave function is represented as in Eqs.~\eqref{general}
\begin{equation}
\Psi(x,y)=\Psi_{\mathrm{inc}}(x,y)+ \int_{0}^{L}  H_0^{(1)}\left (k\sqrt{(x-t)^2+y^2}\, \right )\mu(t)\mathrm{d}t+
\partial_y\int_{0}^{L}  H_0^{(1)}\left (k\sqrt{(x-t)^2+y^2}\,\right )\nu(t)\mathrm{d}t\ .
\end{equation}
 Equations~\eqref{eq_nu} and \eqref{eq_mu} in this case take the form
\begin{eqnarray}
\nu(x)&=&\frac{\tan \pi \alpha }{2} \left [ \Psi_{\mathrm{inc}}(x,0) + \int_0^L H_0^{(1)}(k|x-t|)\mu(t) \mathrm{d}t 
 \right ] \ ,\\
\mu(x)&=& \frac{\tan \pi \alpha}{2}\left [\partial_y \Psi_{\mathrm{inc}}(x,0) - \left ( \frac{\mathrm{d}^2}{\mathrm{d} x^2} +k^2\right ) \int_0^L  H_0^{(1)}(k|x-t|)\nu(t)\mathrm{d}t\right ] \ .
\label{mu_opposite}
\end{eqnarray} 
The main simplification for small distance vortices comes from the fact that when condition 
\begin{equation}
kL\ll 1
\label{kl}
\end{equation}
is fulfilled one can substitute  in the above equations  the asymptotics of the Hankel function at small arguments \cite{bateman}
\begin{equation}
H_0^{(1)}(z)\to \frac{2\mathrm{i}}{\pi}\left ( \ln \frac{x}{2}+\gamma \right )+1 +\mathcal{O}(x^2\ln x)
\label{asymptotic_Hankel}
\end{equation}
where $\gamma=-\Psi(1)$ is the Euler constant and  drop the term proportional to $k^2$ in Eq.~\eqref{mu_opposite}.

After these approximations the equations for  $\nu(x)$ and  $\mu(x)$ become   
\begin{eqnarray}
\nu(x)&=&\frac{\tan \pi \alpha }{2} 
\left [ \Psi_{\mathrm{inc}}(x,0) + \frac{2\mathrm{i}}{\pi}\int_{0}^L \ln \left (\frac{|x-t|}{L} \right )\mu(t) \mathrm{d}t+ \left (\frac{2\mathrm{i}}{\pi}\left ( \ln \frac{kL}{2}+\gamma \right )+1 \right )\int_{0}^L \mu(t) \mathrm{d}t\right ]\ ,
\label{eq_nu_small}\\
\mu(x)&=& \frac{\tan \pi \alpha}{2} \left [ \partial_y \Psi_{\mathrm{inc}}(x,0)  -\frac{2\mathrm{i}}{\pi} \frac{\mathrm{d}^2}{\mathrm{d}x^2}\int_{0}^L\ln \left (\frac{|x-t|}{L} \right ) \nu(t)\mathrm{d}t \right ]\  .
\label{eq_mu_small}
\end{eqnarray}
Deriving Eq.~\eqref{eq_nu_small} on $x$, integrating by part Eq.~\eqref{eq_mu_small}, and taking into account that 
\begin{equation}
\nu(0)=0, \qquad \nu(L)=0
\label{boundary_nu}
\end{equation}
 one  transforms the above equations into the following system of equations
\begin{equation}
\nu^{\prime}(x)=\frac{\tan \pi \alpha }{2}\partial_x\Psi_{\mathrm{inc}}(x,0) +  \frac{\tan \pi \alpha}{\pi \mathrm{i}} \dashint_0^L  \frac{\mu(t)}{t-x}\mathrm{d}t\ ,\qquad 
\mu(x)=\frac{\tan \pi \alpha }{2}\partial_x\Psi_{\mathrm{inc}}(x,0)-\frac{\tan \pi \alpha}{\pi \mathrm{i}} \dashint_{0}^L \frac{\nu^{\prime}(t)}{t-x}\mathrm{d}t \ .
\end{equation}
The equations are decoupled by introducing new variables
\begin{equation}
\zeta_{\pm}(x)= \nu^{\prime}(x)\pm \mathrm{i} \mu(x).
\label{zeta_pm}
\end{equation}
It leads to
\begin{equation}
\zeta_+(x)=\tan \pi \alpha\,  f_+(x)-\frac{\tan \pi \alpha}{\pi }   \dashint_0^L \frac{\zeta_+(t)}{t-x}\mathrm{d}t\ ,\qquad 
\zeta_-(x)=\tan \pi \alpha\,  f_-(x)+ \frac{\tan \pi \alpha}{\pi } \dashint_0^L   \frac{\zeta_-(t)}{t-x}\mathrm{d}t\ ,
\label{xi_eq}
\end{equation}
 where
\begin{equation}
f_{\pm}(x)=\frac{1}{2}(\partial_x\pm\mathrm{i}\partial_y)\Psi_{\mathrm{inc}}(x,0).
\label{f_pm}
\end{equation}  
These equations can be solved by the Rieman-Hilbert  method (see e.g. \cite{singular}). Let us introduce the following functions of complex argument $z$
\begin{equation}
\Phi_{\pm}(z)=\frac{1}{2\pi \mathrm{i}} \int_{0}^L \frac{\zeta_{\pm}(t)}{t-z}\mathrm{d}t .
\label{functions}
\end{equation}
It is plain that 
\begin{equation}
\frac{1}{\pi \mathrm{i}}\dashint_0^L \frac{\zeta_{\pm}(t)}{t-x}\mathrm{d}t=\Phi_{\pm}^{\mathrm{up}}(x)+\Phi_{\pm}^{\mathrm{down}}(x),\qquad 
\zeta_{\pm}(x)=\Phi_{\pm}^{\mathrm{up}}(x)-\Phi_{\pm}^{\mathrm{down}}(x)
\label{zeta}
\end{equation}
where $\Phi_{\pm}^{\mathrm{up}}(x)$ and $\Phi_{\pm}^{\mathrm{down}}(x)$ are the limiting values of functions \eqref {functions} from, respectively, positive and negative $y$.

After a little algebra one gets that Eqs.~\eqref{xi_eq} are equivalent to 
\begin{equation}
\Phi_{\pm}^{\mathrm{up}}(x)=\mathrm{e}^{\mp 2\pi \mathrm{i}\alpha}\Phi_{\pm}^{\mathrm{down}}(x)+\mathrm{e}^{\mp \pi \alpha\mathrm{i}}\sin \pi \alpha \, f_{\pm}(x)
\label{phi_pm}
\end{equation}
whose general solutions are (here $|z|>L$) \cite{singular}
\begin{eqnarray}
\Phi_{+}(z)&=&\frac{C_1}{(z-L)^{\alpha}z^{1-\alpha}} +\frac{\sin \pi \alpha }{2\pi \mathrm{i}(z-L)^{\alpha}z^{1-\alpha}} 
\int_0^L \frac{f_+(t)(L-t)^{\alpha}t^{1-\alpha} }{t-z}\mathrm{d}t\ , \\
\Phi_{-}(z)&=&\frac{C_2}{(z-L)^{1-\alpha}z^{\alpha}}-\frac{\sin \pi \alpha }{2\pi \mathrm{i}(z-L)^{1-\alpha}z^{\alpha}} 
\int_0^L \frac{f_-(t)(L-t)^{1-\alpha}t^{\alpha} }{t-z}\mathrm{d}t 
\nonumber
\label{phi_plus_minus}
\end{eqnarray}
with arbitrary constants $C_1$ and $C_2$.

When condition \eqref{kl} is valid, functions $f_{\pm}(x)$ can be approximated by their values at $0$ and
\begin{eqnarray}
\Phi_{+}(z)&=&\frac{C_1}{(z-L)^{\alpha}z^{1-\alpha}} +\frac{f_+(0)}{2  \mathrm{i}}
\left [1+ \frac{\alpha L-z}{(z-L)^{\alpha}z^{1-\alpha}} \right ] \ , 
\\
\Phi_{-}(z)&=&\frac{C_2}{(z-L)^{1-\alpha}z^{\alpha}}-\frac{f_-(0)}{2 \mathrm{i} } 
\left [1+\frac{(1-\alpha) L-z}{(z-L)^{1-\alpha}z^{\alpha}} \right ] \ .
\nonumber 
\end{eqnarray}
The same expressions can be obtained directly from \eqref{phi_pm} by imposing that $\Phi_{\pm}(z)=\mathcal{O}(z^{-1})$. 

According to Eq.~\eqref{zeta} $\zeta_{\pm}(x)=\Phi_{\pm}^{\mathrm{up}}(x)-\Phi_{\pm}^{\mathrm{down}}(x)$, therefore when $0<x<L$ 
\begin{eqnarray}
\nu^{\prime}(x)&=&-\frac{\mathrm{i}\sin \pi \alpha\,  C_1}{(L-x)^{\alpha}x^{1-\alpha}}-\frac{\mathrm{i}\sin \pi \alpha \, C_2}{(L-x)^{1-\alpha}x^{\alpha}} -\frac{\sin \pi \alpha\, f_+(0)[\alpha L -x]}{2(L-x)^{\alpha}x^{1-\alpha}}+\frac{\sin \pi \alpha\, f_-(0)[(1-\alpha)L -x]}{2(L-x)^{1-\alpha}x^{\alpha}}\ ,\\
\mu(x)&=&-\frac{\sin \pi \alpha\,  C_1 }{(L-x)^{\alpha}x^{1-\alpha}}+\frac{\sin \pi \alpha \, C_2}{(L-x)^{1-\alpha}x^{\alpha}}+
\frac{\mathrm{i}\sin \pi \alpha\, f_+(0)[\alpha L -x]}{2(L-x)^{\alpha}x^{1-\alpha}}+\frac{\mathrm{i}\sin \pi \alpha\, f_-(0)[(1-\alpha)L-x]}{2(L-x)^{1-\alpha}x^{\alpha}}\ .
\nonumber 
\end{eqnarray}
Constants $C_j$ have to be determined from conditions \eqref{boundary_nu}.  First, it is necessary that
\begin{equation}
\int_0^L \nu^{\prime}(x)\mathrm{d}x=0 .
\label{condition_nu}
\end{equation}
Second, from Eq.~\eqref{eq_nu_small} calculated at $x=0$ it follows that   
\begin{equation}
\Psi_{\mathrm{inc}}(0,0)  +\frac{2\mathrm{i}}{\pi}\int_{0}^L \ln\left (\frac{t}{L}\right )\mu(t) \mathrm{d}t+\left (\frac{2\mathrm{i}}{\pi}\left ( \ln \frac{k L}{2}+\gamma \right )+1 \right )\int_{0}^L \mu(t) \mathrm{d}t=0\ .
 \label{condition_mu}
\end{equation}
The Euler integral (see e.g. 1.5 of \cite{bateman_1}) and its derivative
\begin{equation}
\int_0^1 t^{x-1}(1-t)^{y-1} \mathrm{d}t =\frac{\Gamma(x)\Gamma(y)}{\Gamma(x+y)},\qquad 
\int_0^1\ln t\,  t^{x-1}(1-t)^{y-1} \mathrm{d}t =[ \Psi(x)-\Psi(x+y)]\frac{\Gamma(x)\Gamma(y)}{\Gamma(x+y)}
\end{equation}
permit to calculate all necessary integrals analytically. 

From \eqref{condition_nu} it follows that $C_2=-C_1$ and then  Eq.~\eqref{condition_mu} gives
\begin{equation}
 C_1=[\Psi_{\mathrm{inc}}(0,0)+(1-\alpha)f_+(0)L +\alpha f_-(0) L]\Delta,\qquad \Delta= \dfrac{1}{ 2\pi +4\mathrm{i} [\ln (kL/4)+\gamma +\beta(\alpha) }
\label{s_wave}
\end{equation}
with $\gamma=-\Psi(1)$ and 
\begin{equation}
\beta(\alpha) =\ln 2+\frac{1}{2}\Psi(1-\alpha)+\frac{1}{2}\Psi(\alpha)+\gamma .
\end{equation}
The knowledge of $\nu(t)$ and $\nu(t)$ permits to reconstruct the full wave function.  In particular the scattering amplitude at small $kL$ is large only for the $s$-wave scattering and 
\begin{equation}
\mathcal{F}=\int_{0}^L\mu(t)\mathrm{d}t=-2\pi C_1=-\left [1+\frac{2\mathrm{i}}{\pi} \left (\ln \frac{kL}{4} +\gamma +\beta(\alpha) \right ) \right ]^{-1} .
\end{equation}  
In \cite{bms} it has been obtained  in a different manner that 
\begin{equation}
\beta(\alpha) =\ln 4+\frac{1}{2} \Psi\Big (\frac{1-\alpha}{2}\Big )+\frac{1}{2}\Psi\Big (\frac{\alpha}{2}\Big )+\gamma+\frac{\pi}{2\sin \pi \alpha}\ .
\end{equation}
Using Legendre's duplication formula (see e.g. \cite{bateman_1} 1.2.15) it is straightforward to check that  
\begin{equation}
 \Psi\Big (\frac{1-\alpha}{2}\Big )+\Psi\Big (\frac{\alpha}{2}\Big )+\frac{\pi}{\sin \pi \alpha}=\Psi(1-\alpha)+\Psi(\alpha)-2\ln 2\ .
\end{equation}
Therefore these two results are identical. 

The same formulae permit to calculate limiting values of other quantities discussed in the preceding Sections. First one has ($z=x+\mathrm{i}y$, $\bar{z}=x-\mathrm{i}y$)
\begin{eqnarray}
(\partial_x+\mathrm{i}\partial_y)\Psi(x,y)&=&(\partial_x+\mathrm{i}\partial_y) \Psi_{\mathrm{inc}}(x,y) +\frac{2}{\pi} \int_0^L \frac{\zeta_+(t)}{\bar{z}-t}\mathrm{d}t=(\partial_x+\mathrm{i}\partial_y) \Psi_{\mathrm{inc}}(x,y)-4\mathrm{i} \Phi_+(\bar{z})\ ,\\ 
(\partial_x-\mathrm{i}\partial_y)\Psi(x,y)&=&(\partial_x-\mathrm{i}\partial_y) \Psi_{\mathrm{inc}}(x,y)-\frac{2}{\pi} \int_0^L \frac{\zeta_-(t)}{z-t}\mathrm{d}t=(\partial_x-\mathrm{i}\partial_y) \Psi_{\mathrm{inc}}(x,y) +4\mathrm{i} \Phi_-(z) \ .
\nonumber 
\end{eqnarray}
From these expressions it follows (as it has been checked that $\Psi(\vec{L}_j)=0$) that
\begin{equation}
\Psi(x,y)\underset{x\to L_1}{\longrightarrow} a_1 (x-L_1+\mathrm{i}y)^{\alpha} +b_1(x-L_1-\mathrm{i}y)^{1-\alpha},\quad 
\Psi(x,y)\underset{x\to L_2}{\longrightarrow} a_2 (x-L_2+\mathrm{i}y)^{1-\alpha} +b_2(x-L_2-\mathrm{i}y)^{\alpha}
\end{equation}
with
\begin{eqnarray}
a_1&=&- \left [ \frac{ 2\mathrm{i}C_1}{\alpha} -f_-(0)L \right ]  L^{-\alpha} ,\qquad 
b_1=- \left [ \frac{2\mathrm{i}C_1}{1-\alpha} - f_+(0)L \right ]L^{\alpha-1}, \\
a_2&=& \mathrm{e}^{\mathrm{i} \pi \alpha}  \left [ \frac{2\mathrm{i}C_1}{1-\alpha} +f_-(0)L\right ]L^{\alpha-1},\qquad
b_2=-\mathrm{e}^{\mathrm{i}\pi \alpha}\left [ \frac{2\mathrm{i}C_1}{\alpha} +f_+(0)L \right ]L^{-\alpha} .
\nonumber 
\end{eqnarray}
For the Green function one has
\begin{equation}
\Psi_{\mathrm{inc}}(0,0)=\frac{1}{4\mathrm{i}} H^{(1)}_0(k|\vec{x}^{\, \prime}-\vec{L}_2|),\qquad 
(\partial_x\pm \mathrm{i}\partial_y)\Psi_{\mathrm{inc}}(0,0)=\frac{k}{4\mathrm{i}}H^{(1)}_1(k|\vec{x}^{\, \prime}-\vec{L}_2|)\mathrm{e}^{\pm \mathrm{i}\phi^{\prime}}.
\end{equation}
Differentiating these expressions over $L_j$ and using definitions \eqref{dif_a_b_two_vortices} one finds that 
\begin{equation} 
\left (\begin{array}{cc}\beta_2 & \epsilon_2\\ \delta_2 & \zeta_2 \end{array}\right) \underset{L\to 0}{\longrightarrow }
\left (\begin{array}{lr}  
 \alpha \left [1+\dfrac{2\mathrm{i}\Delta}{1-\alpha}\right ]L^{2\alpha-2}  & 
 2\mathrm{i}\Delta  L^{-1} \\
-2\mathrm{i}\Delta  L^{-1}&
-(1-\alpha) \left [1+\dfrac{2\mathrm{i}\Delta}{\alpha}\right ]L^{-2\alpha}
 \end{array}\right )\mathrm{e}^{\mathrm{i}\pi \alpha}
 \label{second_limit}
\end{equation} 
and 
\begin{equation} 
\left (\begin{array}{cc}\beta_1 & \epsilon_1\\ \delta_1 & \zeta_1 \end{array}\right) \underset{L\to 0}{\longrightarrow }
\left (\begin{array}{lr}  
- (1-\alpha) \left [1+\dfrac{2\mathrm{i}\Delta}{\alpha}\right ]L^{-2\alpha}  & 
 2\mathrm{i}\Delta  L^{-1} \\
- 2\mathrm{i}\Delta  L^{-1}&
\alpha \left [1+\dfrac{2\mathrm{i}\Delta}{1-\alpha}\right ] L^{2\alpha-2} 
 \end{array}\right )\mathrm{e}^{-\mathrm{i}\pi \alpha}\ .
 \label{first_limit}
\end{equation}
Limiting behaviors of coefficients $g_j$ and $f_j$ follow from Eqs.~\eqref{translation_1}  and \eqref{translation_2}
\begin{equation}
g_2\underset{L\to 0}{\longrightarrow } \frac{4\mathrm{i} \alpha}{1-\alpha}\Delta L^{2\alpha-2},\quad g_3\underset{L\to 0}{\longrightarrow } \frac{4\mathrm{i}(1- \alpha)}{\alpha}\Delta  L^{-2\alpha},\quad 
f_2\underset{L\to 0}{\longrightarrow } \frac{4\mathrm{i} (1-\alpha)}{\alpha}\Delta L^{-2\alpha},\quad f_3\underset{L\to 0}{\longrightarrow } \frac{4\mathrm{i} \alpha}{1-\alpha}\Delta  L^{2\alpha-2}.
\label{L_to_zero}
\end{equation}


\subsection{Two vortices with arbitrary fluxes at small distances}

General case consists of two vortices with fluxes $\alpha_1$ and $\alpha_2$ ($0<\alpha_j<1$) separated by a distance, $L$,  obeying \eqref{kl}.  The principal difference with opposite flux vortices (i.e. with $\alpha_2=1-\alpha_1$)  considered above is the existence of an additional cut going from infinity to the vortex positions. For convenience we choose the both  cuts along the $x$-axis as in Fig.~ref{contour}  such that the function $\chi(x)$ is as in Eq.~\eqref{chi}. The reflected field is chosen as in Eq.~\eqref{general} which leads to Eqs.~\eqref{eq_nu} and \eqref{eq_mu} for unknown functions $\nu(x)$ and $\mu(x)$. As a consequence, one has to know these functions along the whole negative $x$-axis and not only at short cut between two vortices. To take into account the condition \eqref{kl} explicitly it is convenient to look for the wave function of this problem in the form slightly different from Eq.~\eqref{general}, namely 
\begin{equation}
\Psi(\vec{x}\, )=\Psi_{\beta}(\vec{x}\, )+ \int_{-\infty}^{L}  H_0^{(1)}\left (k\sqrt{(x-t)^2+y^2}\, \right )\mu(t)\mathrm{d}t+
\partial_y\int_{-\infty}^{L}  H_0^{(1)}\left (k\sqrt{(x-t)^2+y^2}\,\right )\nu(t)\mathrm{d}t \ .
\label{different}
\end{equation}
Here $\Psi_{\beta}(\vec{x}\, )$ is the one-vortex solution generated by the desired incident field $\Psi^{\mathrm{inc}}(\vec{x}\, ) $ (see Appendix~\ref{one_vortex}) multiplied by $\mathrm{e}^{\mathrm{i}\pi \alpha_1}$ for the cuts as in Fig.~\ref{contour}.  It corresponds to one vortex with flux equals the total flux of two vortices  
\begin{equation}
\beta=\{\alpha_1+\alpha_2\}=\alpha_1+\alpha_2-\eta, \qquad \eta=\left \{ \begin{array}{cc} 0,& 0<\alpha_1+\alpha_2<1\\
1,& 1<\alpha_1+\alpha_2<2\end{array}\right .
\label{beta}
\end{equation}
situated at point $L_2=0$. 

Functions $\nu(x)$ and $\mu(x)$ have to fulfill equations \eqref{eq_nu} and \eqref{eq_mu} which we rewrite below for the convenience
\begin{eqnarray}
\nu(x)&=& \tfrac{1}{2}\tan \pi \chi(x)  \int_{-\infty}^L  H_0^{(1)}(k|x-t|)\mu(t) \mathrm{d}t + \mathcal{F}(x,0)\ , \\
\mu(x) &=& -\tfrac{1}{2}\tan \pi \chi(x) \left ( \partial_x^2 +k^2\right ) \int_{-\infty}^L   H_0^{(1)}(k|x-t|)\nu (t)\mathrm{d}t +\partial_y \mathcal{F}(x,0)\ .
\end{eqnarray}
As function  $\Psi_{\beta}(\vec{x}\, )$ obeys the correct boundary conditions at the cut $(-\infty,0]$,  functions $\mathcal{F}(x,0)$ and $\partial_y \mathcal{F}(x,0)$ have the form 
\begin{equation}
\mathcal{F}(x,0)=\Theta(x)\frac{\tan \pi \alpha_1}{2}\Psi_{\beta}(x,0 ),\qquad  \partial_y \mathcal{F}(x,0)=\Theta(x)\frac{\tan \pi \alpha_1}{2}\partial_y \Psi_{\beta}(x,0 )
\end{equation}
where $\Theta(x)=0$ for $x<0$ and  $\Theta(x)=1$ for $x>0$.  

As the vortex separation is assumed to be  small (cf. Eq.\eqref{kl}), functions $\nu(x)$ and $\mu(x)$ in \eqref{different} should decrease quickly from vortex positions so that all integrals are dominated by a vicinity of the origin. 

In such conditions one can (i) approximate the above equations using \eqref{asymptotic_Hankel} as it has been done in the previous Section and (ii) use the small-$\vec{x}$ asymptotics of function   $\Psi_{\beta}(\vec{x}\, )$ given by \eqref{asymptotic_j} 
\begin{equation}
\Psi_{\beta}(\vec{x}\, )\underset{\vec{x}\to 0}{\longrightarrow} a(x+\mathrm{i}y)^{\beta}+b (x-\mathrm{i}y)^{1-\beta}
\label{psi_beta}
\end{equation}
with certain (known) quantities $a$ and $b$ (fixed by the quantity considered). For the Green functions this expansion is given by Eq.~\eqref{one_vortex_asymptotic} 
\begin{equation}
a\equiv a(\vec{x}^{\, \prime})=
  -\frac{\mathrm{i}\, k^{\beta}\mathrm{e}^{\mathrm{i}\pi \alpha_1}}{  2^{\beta+2}  \Gamma(1+\beta)}  H_{\beta}^{(1)}(kR)\mathrm{e}^{-\mathrm{i}\beta \phi},\qquad 
b\equiv b(\vec{x}^{\, \prime})=
-\frac{\mathrm{i}\, k^{1-\beta}\mathrm{e}^{\mathrm{i}\pi \alpha_1}}{  2^{3-\beta}  \Gamma(2-\beta)} H_{1-\beta}^{(1)}(kR)\mathrm{e}^{\mathrm{i}(1-\beta) \phi} .
\end{equation}
In the small-distance approximation Eqs.~\eqref{eq_nu} and \eqref{eq_mu}) take the form (for  $-\infty<x<L$)
\begin{eqnarray}
\nu(x)&=&\frac{\mathrm{i} \tan \pi \chi(x)}{\pi} 
\left [ \int_{-\infty}^L  \ln\left (\frac{|x-t|}{L}\right )\mu(t) \mathrm{d}t+\left (\ln \frac{k L}{2}+\gamma  + \frac{\pi}{2\mathrm{i}} \right )\int_{-\infty}^L  \mu(t) \mathrm{d}t\right ] +\mathcal{F}(x)\ ,
\\
\mu(x)&=& -\frac{\mathrm{i} \tan \pi \chi(x)}{\pi}\left [ \frac{\mathrm{d}^2}{\mathrm{d}x^2} \int_{-\infty}^L \ln(|x-t|)\nu(t)\mathrm{d}t \right ]+\partial_y \mathcal{F}(x)
\end{eqnarray}
with $\chi(x)$ defined in Eq.~\eqref{chi}. 

In order that the contributions from large negative values of $t$ will be small,  the following asymptotics is required
\begin{equation}
\mu(t)\underset{t\to -\infty}{\sim} |t|^{-\gamma_1},\qquad \nu^{\prime}(t)\underset{t\to -\infty}{\sim }|t|^{-\gamma_2}, \qquad \gamma_j >1.
\label{negative_x}
\end{equation}
Differentiating Eq.~\eqref{eq_nu_small} and introducing functions \eqref{zeta_pm} one gets  the equations
\begin{eqnarray}
\zeta_+(x)&=& \tan \pi \alpha_1\, \Theta(x) f_+(x) - \frac{\tan \pi \chi(x)}{\pi}\dashint_{-\infty}^L  \frac{\zeta_+(t)}{t-x} \mathrm{d}t \ ,\label{f}\\
\zeta_-(x)&=& \tan \pi \alpha_1\, \Theta(x) f_-(x)+
 \frac{\tan \pi \chi(x)}{\pi}\dashint_{\infty}^L   \frac{\zeta_-(t)}{t-x} \mathrm{d}t\ . 
\label{g}
\end{eqnarray}
For the incident field \eqref{psi_beta}
\begin{equation}
f_+(x)\equiv \tfrac{1}{2}(\partial_x+\mathrm{i}\partial_y) \Psi_{\beta}(x,0)=(1-\beta)b x^{-\beta},\qquad 
f_-(x)\equiv \tfrac{1}{2}(\partial_x-\mathrm{i}\partial_y) \Psi_{\beta}(x,0) =\beta a x^{\beta-1}.
\end{equation} 
Introducing similar to  Eq.~\eqref{functions} analytic functions
\begin{equation}
\Phi_{\pm}(z)=\frac{1}{2\pi \mathrm{i}} \int_{-\infty}^L \frac{\zeta_{\pm}(t)}{t-z}\mathrm{d}t
\end{equation}
permits to find the general solution of Eqs.~\eqref{f} and \eqref{g} (cf. Eqs.~\eqref{phi_plus_minus})
\begin{eqnarray}
\Phi_+(z)&=&\frac{C_1}{z^{\alpha_2}(z-L)^{\alpha_1}}+\frac{\sin \pi \alpha_1}{2\pi \mathrm{i} \, z^{\alpha_2}(z-L)^{\alpha_1}}\int_0^L 
\frac{t^{\alpha_2}(L-t)^{\alpha_1}}{t-z}\, f_+(t)\, \mathrm{d}t \ ,\label{phi_1}\\
\Phi_-(z)&=&\frac{C_0}{z^{1-\alpha_2}(z-L)^{1-\alpha_1}}-\frac{\sin \pi \alpha_1}{2\pi \mathrm{i}\,  z^{1-\alpha_2}(z-L)^{1-\alpha_1}}\int_0^L \frac{t^{1-\alpha_2}(L-t)^{1-\alpha_1}}{t-z}\, f_-(t)\, \mathrm{d}t\ .
\label{phi_2}
\end{eqnarray}
Branches are fixed by requiring that fractional powers are real at real $z>L$. Imposing the correct behavior at large negative $x$ \eqref{negative_x}, one concludes that  for  $0<\alpha_1+\alpha_2<1$ (i.e. $\eta=0$) constant $C_1=0$ and for  $0<\alpha_1+\alpha_2<1$ (i.e. $\eta=1$) constant $C_0=0$. Notice that for opposite fluxes (i.e.  when $\alpha_2=1-\alpha_1$) the both constants are non-zero.    

The remaining integrals in Eqs.~\eqref{phi_1} and \eqref{phi_2} reduce to the following ones
\begin{equation}
\int_0^L \frac{t^{\alpha} (L-t)^{1-\alpha}}{t-z}\mathrm{d}t=\frac{\pi}{\sin \pi \alpha}\left [ z^{\alpha}(z-L)^{1-\alpha}-z+(1-\alpha)L\right ],\quad
\int_0^L \frac{t^{-\alpha} (L-t)^{\alpha}}{t-z}\mathrm{d}t=\frac{\pi}{\sin \pi \alpha}\left [ z^{-\alpha}(z-L)^{\alpha}-1 \right ] .
\end{equation}
For $\eta=0$ one gets     
\begin{eqnarray}
\Phi_+^{(0)}(z)&=& \frac{(1-\alpha_1-\alpha_2)b}{2\mathrm{i}z^{\alpha_2}(z-L)^{\alpha_1}}\left [z^{-\alpha_1}(z-L)^{\alpha_1}-1 \right ] \ ,  \\
\Phi_-^{(0)}(z)&=&\frac{C_0}{z^{1-\alpha_2}(z-L)^{1-\alpha_1}}-\frac{(\alpha_1+\alpha_2)a }{2 \mathrm{i}\,  z^{1-\alpha_2}(z-L)^{1-\alpha_1}}
\left [z^{\alpha_1}(z-L)^{1-\alpha_1}-z+(1-\alpha_1)L \right ]\ . 
\nonumber
\end{eqnarray}
For  $\eta=1$ 
\begin{eqnarray}
\Phi_+^{(1)}(z)&=& \frac{C_1}{z^{\alpha_2}(z-L)^{\alpha_1}} +\frac{(2-\alpha_1-\alpha_2)b}{2 \mathrm{i}\,  z^{\alpha_2}(z-L)^{\alpha_1}}\left [z^{1-\alpha_1}(z-L)^{\alpha_1}-z+\alpha_1 L  \right ] \ ,\nonumber \\
\Phi_-^{(1)}(z)&=&-\frac{(\alpha_1+\alpha_2-1) a}{2 \mathrm{i}\,  z^{1-\alpha_2}(z-L)^{1-\alpha_1}}
\left [ z^{\alpha_1-1}(z-L)^{1-\alpha_1}-1  \right ]\ .
\end{eqnarray}
Functions $\zeta_{\pm}^{(\eta)}(t)$ are boundary jumps of these functions (cf. \eqref{zeta}). They have different forms depending on the cuts  
\begin{equation}
\zeta_{\pm}^{(\eta)}(t)=\left \{ \begin{array}{cc} 
\sin \pi \alpha_1\,  F_{\pm}^{(\eta)}(t) ,  & 0<t<L\\
\sin \pi (\alpha_1+\alpha_2)\,   G_{\pm}^{(\eta)}(t)  , & t<0 \end{array}\right .\ ,
\end{equation}
where 
\begin{itemize}
\item $\eta=0$
\begin{eqnarray}
F_+^{(0)}(t)&=& \frac{(1-\alpha_1-\alpha_2)b }{t^{\alpha_2}(L-t)^{\alpha_1}} , \quad
G_+^{(0)}(t)=-(1-\alpha_1-\alpha_2)b \,  \frac{(-t)^{-\alpha_1}(L-t)^{\alpha_1}-1}{(-t)^{\alpha_2}(L-t)^{\alpha_1}},\\
F_-^{(0)}(t)&=& -\frac{2\mathrm{i}C_0 +(\alpha_1+\alpha_2)a [t-(1-\alpha_1)L ] }{t^{1-\alpha_2}(L-t)^{1-\alpha_1}} ,\quad 
G_-^{(0)}(t)= \frac{ 2\mathrm{i}C_0 +(\alpha_1+\alpha_2)a [ (-t)^{\alpha_1} (L-t)^{1-\alpha_1}+t-(1-\alpha_1)L ]}{(-t)^{1-\alpha_2}(L-t)^{1-\alpha_1}}.
\nonumber
\end{eqnarray}
\item $\eta=1$
\begin{eqnarray}
F_+^{(1)}(t)&=& -\frac{2\mathrm{i}C_1-(2-\alpha_1-\alpha_2)b [t-\alpha_1 L]}{t^{\alpha_2}(L-t)^{\alpha_1}} , \quad 
G_+^{(1)}(t)= - \frac{ 2\mathrm{i}C_1 - (2-\alpha_1-\alpha_2)b  [ (-t)^{1-\alpha_1}(L-t)^{\alpha_1} +t-\alpha_1 L] }{(-t)^{\alpha_2}(L-t)^{\alpha_1}}, \nonumber \\
F_-^{(1)}(t)&=& -\frac{(\alpha_1+\alpha_2-1)a }{t^{1-\alpha_2}(L-t)^{1-\alpha_1}} ,\quad
G_-^{(1)}(t)= -\frac{(\alpha_1+\alpha_2-1)a [ (-t)^{\alpha_1-1} (L-t)^{1-\alpha_1}-1]}{(-t)^{1-\alpha_2}(L-t)^{1-\alpha_1}}.
\end{eqnarray}
\end{itemize}
Functions $\mu(t)$ and $\nu^{\prime}(t)$ are 
\begin{equation}
\mu^{(\eta)}(t)=\frac{\mathrm{i}}{2}(\zeta_-^{(\eta)}(t)-\zeta_+^{(\eta)}(t)),\qquad \nu^{(\eta)\prime}(t)=\frac{1}{2}(\zeta_-^{(\eta)}(t)+\zeta_+^{(\eta)}(t)).
\end{equation}
For all values of $\eta$ functions $\Phi_{\pm}(z)$ have singularities at $z=0$ and $z=1$ and decay as $z^{-\gamma}$ with $\gamma>1$ at infinity.

Therefore (which can also be checked by direct calculations)
\begin{equation}
\int_{-\infty}^L\zeta_{\pm}^{(\eta)}\mathrm{d}t=0, 
\end{equation}
and the only condition to fulfill is (as $\mathcal{F}(0)=0$)
\begin{equation}
\int_{-\infty}^L \ln\left ( \frac{|t|}{L}\right )\mu(t)\mathrm{d}t=0 .
\label{last_condition}
\end{equation}
The necessary integrals can be calculated by differentiation of the Euler integral and the answer is
\begin{eqnarray}
\int_{-\infty}^L \zeta_-^{(0)}(t)\ln\left ( \frac{|t|}{L}\right )\mathrm{d}t &=&
-Z L^{\alpha_1+\alpha_2-1} \frac{\Gamma(\alpha_1) \Gamma(\alpha_2)}{\Gamma(\alpha_1+\alpha_2)} \left [ 2\mathrm{i}C_0+ \alpha_1(\alpha_1 +\alpha_2-1)a \right ]\ ,\nonumber \\
\int_{-\infty}^L \zeta_+^{(0)}(t)\ln\left ( \frac{|t|}{L}\right )\mathrm{d}t&=&
-Z L^{1-\alpha_1-\alpha_2}b \, \frac{\Gamma(1-\alpha_1) \Gamma(1-\alpha_2)}{\Gamma(2-\alpha_1-\alpha_2)}\ , \\
\int_{-\infty}^L \zeta_-^{(1)}(t)\ln\left ( \frac{|t|}{L}\right )\mathrm{d}t&=&
-Z L^{\alpha_1+\alpha_2-1}a \, \frac{\Gamma(\alpha_1) \Gamma(\alpha_2)}{\Gamma(\alpha_1+\alpha_2-1)}\ , \nonumber\\
\int_{-\infty}^L \zeta_+^{(1)}(t)\ln\left ( \frac{|t|}{L}\right )\mathrm{d}t&=&
Z L^{1-\alpha_1-\alpha_2} \frac{\Gamma(1-\alpha_1) \Gamma(1-\alpha_2)}{\Gamma(2-\alpha_1-\alpha_2)} \left [ 2\mathrm{i}C_1- (1-\alpha_1)(1-\alpha_1 -\alpha_2)b \right ]\ ,
\nonumber
\end{eqnarray}
where
\begin{equation}
Z=\frac{\pi \cos(\pi(\alpha_1+\alpha_2))\sin \pi \alpha_1 }{\sin (\pi(\alpha_1+\alpha_2))}\ .
\end{equation}
Condition \eqref{last_condition} fixes values of constants $C_{\eta}$
\begin{eqnarray}
2\mathrm{i} C_0&=&\alpha_1(1-\alpha_1-\alpha_2)a\,  L +Rb\,   L^{2(1-\alpha_1-\alpha_2) },
\label{C_zero}\\
2\mathrm{i} C_1&=&-(1-\alpha_1)(\alpha_1+\alpha_2-1)b \,  L- Ta\, L^{2(\alpha_1+\alpha_2-1) }, 
\label{C_one}
\end{eqnarray}
where 
\begin{equation}
R\equiv R(\alpha_1,\alpha_2)=\frac{\Gamma(1-\alpha_1) \Gamma(1-\alpha_2)\Gamma(\alpha_1+\alpha_2) }{\Gamma(\alpha_1) \Gamma(\alpha_2)\Gamma(1-\alpha_1-\alpha_2)},\qquad 
T=R(1-\alpha_1,1-\alpha_2).
\label{R_T}
\end{equation}
Calculating the derivatives one gets the following relations
\begin{equation}
\tfrac{1}{2}(\partial_x+\mathrm{i}\partial_y)\Psi(x,y)=\tfrac{1}{2}(\partial_x+\mathrm{i}\partial_y)\Psi_{\beta}(x,y)-2\mathrm{i}\Phi_+(\bar{z}),\quad \tfrac{1}{2}(\partial_x-\mathrm{i}\partial_y)\Psi(x,y)=\tfrac{1}{2}(\partial_x-\mathrm{i}\partial_y)\Psi_{\beta}(x,y)+2\mathrm{i}\Phi_-(\bar{z}).
\end{equation} 
From the above formulae it follows that in a small vicinity of the origin
\begin{eqnarray}
\tfrac{1}{2}(\partial_x+\mathrm{i}\partial_y)\Psi^{(0)}(x,y)&=&\frac{(1-\alpha_1-\alpha_2)b }{\bar{z}^{\alpha_2}(\bar{z}-L)^{\alpha_1} },\quad 
\tfrac{1}{2}(\partial_x-\mathrm{i}\partial_y)\Psi^{(0)}(x,y)=\frac{2 \mathrm{i} C_0+(\alpha_1+\alpha_2)a (z-(1-\alpha_1)L) }{z^{1-\alpha_2}(z-L)^{1-\alpha_1} },\\
\tfrac{1}{2}(\partial_x+\mathrm{i}\partial_y)\Psi^{(1)}(x,y)&=&-\frac{2 \mathrm{i} C_1-(2-\alpha_1-\alpha_2)b (z-\alpha_1 L )}{\bar{z}^{\alpha_2}(\bar{z}-L)^{\alpha_1} }, \quad
\tfrac{1}{2}(\partial_x-\mathrm{i}\partial_y)\Psi^{(1)}(x,y)=\frac{(\alpha_1+\alpha_2-1)a }{z^{1-\alpha_2}(z-L)^{1-\alpha_1} }.
\nonumber
\end{eqnarray}
From these expressions it is possible to calculate functions $a_j$ and $b_j$ from the definition \eqref{x_Lj_alpha_j} and  Eqs.~\eqref{C_zero} and \eqref{C_one}   
\begin{eqnarray}
a_1^{(0)}&=&\frac{2\mathrm{i}C_0+\alpha_1(\alpha_1+\alpha_2)a\, L}{\alpha_1L^{1-\alpha_2}}= aL^{\alpha_2}+\frac{R b}{\alpha_1} L^{1-2\alpha_1-\alpha_2},\quad
b_1^{(0)}= \frac{(1-\alpha_1-\alpha_2)b}{(1-\alpha_1)L^{\alpha_2}}\\
a_2^{(0)}&=&-\mathrm{e}^{\mathrm{i}\pi \alpha_1}\frac{2\mathrm{i}C_0-(1-\alpha_1)(\alpha_1+\alpha_2)a\, L }{\alpha_2 L^{1-\alpha_1}}=
\mathrm{e}^{\mathrm{i}\pi \alpha_1} \left [ a L^{\alpha_1} -\frac{R b}{\alpha_2} L^{1-2\alpha_2-\alpha_1}\right ], \quad  b_2^{(0)}=\mathrm{e}^{\mathrm{i}\pi \alpha_1} \frac{(1-\alpha_1-\alpha_2)b}{(1-\alpha_2)L^{\alpha_1}},
\nonumber
\end{eqnarray}
and 
\begin{eqnarray}
a_1^{(1)}&=&\frac{(\alpha_1+\alpha_2-1)a }{\alpha_1 L^{1-\alpha_2}},\quad 
b_1^{(1)}=\frac{-2\mathrm{i}C_1 +(1-\alpha_1)(2-\alpha_1-\alpha_2) b }{(1-\alpha_1)L^{\alpha_2}}= b L^{1-\alpha_2}+\frac{T a }{1-\alpha_1}  L^{2\alpha_1+\alpha_2-2},\\  
a_2^{(1)}&=&-\mathrm{e}^{\mathrm{i}\pi \alpha_1} \frac{(\alpha_1+\alpha_2-1)a }{\alpha_2 L^{1-\alpha_1}},\quad
b_2^{(1)}=-\mathrm{e}^{\mathrm{i}\pi \alpha_1}  \frac{2\mathrm{i}C_1  +\alpha_1 (2-\alpha_1-\alpha_2)b\, L}{(1-\alpha_2)L^{\alpha_1}}=
\mathrm{e}^{\mathrm{i}\pi \alpha_1}\left [ -bL^{1-\alpha_1}+\frac{T a}{1-\alpha_2} L^{\alpha_1+2\alpha_2-2}\right ].
\nonumber
\end{eqnarray}
Differentiating these expressions on $L_j$, taking into account that derivatives of $a$ and $b$ will contain additional smallness and, therefore, can be considered as constants, one finds
\begin{itemize}
\item $\eta=0$
 \begin{equation} 
\left (\begin{array}{cc}\beta_1 & \epsilon_1\\ \delta_1 & \zeta_1 \end{array}\right) \underset{L\to 0}{\longrightarrow }
\left (\begin{array}{cc}  
- \alpha_2  L^{\alpha_2-\alpha_1-1 }  & 
 -\dfrac{1-\alpha_2}{\alpha_1} R\,  L^{-\alpha_1-\alpha_2 } \\
0 &
\dfrac{\alpha_2(1-\alpha_2)}{1-\alpha_1} L^{\alpha_1-\alpha_2-1} 
 \end{array}\right )\mathrm{e}^{-\mathrm{i}\pi \alpha_1}\ ,
 \label{first_L_to_zero_0}
\end{equation}
 \begin{equation} 
\left (\begin{array}{cc}\beta_2 & \epsilon_2\\ \delta_2 & \zeta_2 \end{array}\right) \underset{L\to 0}{\longrightarrow }
\left (\begin{array}{cc}  
 \alpha_1  L^{\alpha_1-\alpha_2-1 }  & 
 -\dfrac{1-\alpha_1}{\alpha_2} R\,  L^{-\alpha_1-\alpha_2 } \\
0 &
-\dfrac{\alpha_1(1-\alpha_1)}{1-\alpha_2} L^{\alpha_2-\alpha_1-1} 
 \end{array}\right )\mathrm{e}^{\mathrm{i}\pi \alpha_1}\ .
 \label{second_L_to_zero_0}
\end{equation}
\item $\eta=1$
 \begin{equation} 
\left (\begin{array}{cc}\beta_1 & \epsilon_1\\ \delta_1 & \zeta_1 \end{array}\right) \underset{L\to 0}{\longrightarrow }
\left (\begin{array}{cc}  
- \dfrac{\alpha_2(1-\alpha_2)}{\alpha_1} L^{\alpha_2-\alpha_1-1 }  &  0\\
 \dfrac{\alpha_2}{1-\alpha_1} T \,  L^{\alpha_1+\alpha_2-2 }  &
(1-\alpha_2) L^{\alpha_1-\alpha_2-1} 
 \end{array}\right )\mathrm{e}^{-\mathrm{i}\pi \alpha_1}\ ,
 \label{first_L_to_zero_1}
\end{equation}
 \begin{equation} 
\left (\begin{array}{cc}\beta_2 & \epsilon_2\\ \delta_2 & \zeta_2 \end{array}\right) \underset{L\to 0}{\longrightarrow }
\left (\begin{array}{cc}  
 \dfrac{\alpha_1(1-\alpha_1)}{\alpha_2}  L^{\alpha_1-\alpha_2-1 }  & 0\\ 
 \dfrac{\alpha_1}{1-\alpha_2} T\,  L^{\alpha_1+\alpha_2-2 }  &
-(1-\alpha_1) L^{\alpha_2-\alpha_1-1} 
 \end{array}\right )\mathrm{e}^{\mathrm{i}\pi \alpha_1}\ .
 \label{second_L_to_zero_1}
\end{equation} 
\end{itemize} 
In all cases relation \eqref{relation_2_1} is fulfilled with 
\begin{equation}
\rho=\frac{\alpha_1(1-\alpha_1)}{\alpha_2(1-\alpha_2)}\mathrm{e}^{2\pi \mathrm{i}\alpha_1}\ .
\end{equation}
Coefficients $\delta_1$ and $\delta_2$ for $\eta=0$ and $\epsilon_1$ and $\epsilon_2$ for $\eta=1$ are zero in the leading order. Instead the explicit calculation of the next terms it is convenient to use Eqs.~\eqref{translation_final_ab}  by taking into account that
\begin{equation}
(\partial_x+\mathrm{i} \partial_y)a=r_1 b,\qquad 
r_1=-k^{2\beta}2^{1-2\beta}\mathrm{e}^{-\mathrm{i}\pi \beta}\frac{\Gamma(2-\beta)}{\Gamma(1+\beta)},\qquad
(\partial_x-\mathrm{i} \partial_y)b=r_2 a,\qquad 
r_2=-\frac{k^2}{r_1}.
\end{equation}
Comparing coefficients in front of $a$ and $b$ one gets the following limiting values
\begin{itemize}
\item $\eta=0$
\begin{equation}
\delta_1=-\frac{\alpha_2(1-\alpha_1-\alpha_2)}{2(1-\alpha_1)(\alpha_1+\alpha_2)}r_2 \mathrm{e}^{-\mathrm{i}\pi \alpha_1}L^{-\alpha_1-\alpha_2},\qquad 
\delta_2=-\frac{\alpha_1(1-\alpha_1-\alpha_2)}{2(1-\alpha_2)(\alpha_1+\alpha_2)}r_2 \mathrm{e}^{\mathrm{i}\pi \alpha_1}L^{-\alpha_1-\alpha_2},
\label{additional_L_to_zero_0}
\end{equation}
and 
\begin{eqnarray}
g_2&=&\frac{\alpha_1(1-\alpha_1-\alpha_2)}{2(1-\alpha_1)(\alpha_1+\alpha_2)}r_2 L^{-2\alpha_2},\qquad 
g_3=-\frac{2(1-\alpha_1)}{\alpha_1}R L^{-2\alpha_1}+\frac{1-\alpha_1}{1-\alpha_1-\alpha_2}r_1 L^{2\alpha_2},\\ 
f_2&=&\frac{\alpha_2(1-\alpha_1-\alpha_2)}{2(1-\alpha_2)(\alpha_1+\alpha_2)}r_2 L^{-2\alpha_1},\qquad
f_3=-\frac{2(1-\alpha_2)}{\alpha_2}R L^{-2\alpha_1}+\frac{1-\alpha_2}{1-\alpha_1-\alpha_2}r_1 L^{2\alpha_1}.
\nonumber
\end{eqnarray}
\item $\eta=1$
\begin{equation}
\epsilon_1=\frac{(1-\alpha_2)(\alpha_1+\alpha_2-1)}{2\alpha_1 (2-\alpha_1-\alpha_2)}r_1 \mathrm{e}^{-\mathrm{i}\pi \alpha_1}L^{\alpha_1+\alpha_2-2},\qquad 
\epsilon_2=\frac{(1-\alpha_1)(\alpha_1+\alpha_2-1)}{2\alpha_2(2-\alpha_1-\alpha_2)}r_1 \mathrm{e}^{\mathrm{i}\pi \alpha_1}L^{\alpha_1+\alpha_2-2}, 
\end{equation}
and 
\begin{eqnarray}
g_2&=&-\frac{2\alpha_1}{1-\alpha_1}T L^{2\alpha_1-2}+   \frac{\alpha_1}{\alpha_1+\alpha_2-1} r_2 L^{2-2\alpha_2},\qquad 
g_3=\frac{(1-\alpha_1)(\alpha_1+\alpha_2-1)}{\alpha_1(2-\alpha_1-\alpha_2)}r_1 L^{2\alpha_2-2}, \\ 
f_2&=&-\frac{2\alpha_2}{1-\alpha_2}T L^{2\alpha_2-2}+   \frac{\alpha_2}{\alpha_1+\alpha_2-1} r_2 L^{2-2\alpha_1},\qquad 
f_3=\frac{(1-\alpha_2)(\alpha_1+\alpha_2-1)}{\alpha_2(2-\alpha_1-\alpha_2)}r_1 L^{2\alpha_1-2} .
\nonumber
\end{eqnarray}
\end{itemize}
In these calculations it has been taken into account  that $(\partial_x-\mathrm{i} \partial_y)a$ and $(\partial_x+\mathrm{i} \partial_y)b$ correspond to higher order terms in expansion \eqref{psi_beta}, therefore they were put to zero in the leading order. 

The small-distance behavior of $y=\delta_1\epsilon_2$ follows directly from the above expressions. For $\eta=0$
\begin{equation}
y\underset{L\to 0}{\longrightarrow} \left ( \frac{k}{2}\right )^2 \left (\frac{k L}{2}\right )^{-2(\alpha_1+\alpha_2)}
\frac{\sin^2 \pi(\alpha_1+\alpha_2)}{\sin \pi \alpha_1 \sin \pi \alpha_2}\frac{\Gamma^4(\alpha_1+\alpha_2)}{\Gamma^2(\alpha_1)\Gamma^2(\alpha_2)}\mathrm{e}^{\mathrm{i}\pi(\alpha_1+\alpha_2) }\ .
\label{y_L_to_zero_0}
\end{equation}
For $\eta=1$ the limiting behavior of $y$ is given by the same formula but with substitution $\alpha_1\to 1-\alpha_1$, $\alpha_2\to 1-\alpha_2$. 


\subsection{Two vortices at large distances}

The knowledge of the one-vortex solution (see Appendix~\ref{one_vortex}) permits to calculate the two-vortex case within the perturbation series when the distance between vortices is large. Consider functions $A_1(\vec{x}\,)$ and $B_1(\vec{x}\,)$. In the lowest order, when the second vortex is absent, their  asymptotics  is determined  by Eqs.~\eqref{F_j_G_j} and \eqref{zeroth}. The existence of the second vortex even at very large distance modifies  these expressions to the following ones 
\begin{equation}
A_1(\vec{x}_1)\underset{|\vec{x}_1|\to\infty }{\longrightarrow} \sqrt{\frac{2}{\pi\mathrm{i} k r}} \mathrm{e}^{\mathrm{i}kr} f_1^{(0)}(\theta_1) \mathrm{e}^{\mathrm{i}\alpha_1 \theta_1+\mathrm{i}\alpha_2 \theta_2},\qquad
B_1(\vec{x}_1)\underset{|\vec{x}_1|\to\infty }{\longrightarrow} \sqrt{\frac{2}{\pi\mathrm{i} k r}} \mathrm{e}^{\mathrm{i}kr} g_1^{(0)}(\theta_1) \mathrm{e}^{\mathrm{i}\alpha_1 \theta_1+\mathrm{i}\alpha_2 \theta_2} .
\end{equation}
Here $\theta_1$ is the polar angle around the first vortex, $\vec{x}_1=(r\cos \theta_1, r\sin \theta_1)$, and $\theta_2$ indicates the polar angle with the center in the second vortex, $\vec{x}_2=(R\cos \theta_2, R\sin \theta_2)$. It is assumed  that vortices and cuts  are such  as indicated in Fig.~\ref{contour}. Functions $f_j^{(0)}(\theta)$ and $g_j^{(0)}(\theta)$ with $j=1,2$  are obtained from Eq.~\eqref{zeroth} by extracting the factor $\mathrm{e}^{\mathrm{i}\alpha \theta} $
\begin{equation}
 f_j^{(0)}(\theta)=-\mathrm{i} c_3(\alpha_j) \mathrm{e}^{-\mathrm{i}\theta +\mathrm{i}\pi\alpha_j/2},\qquad 
 g_j^{(0)}(\theta)=c_4(\alpha_j) \mathrm{e}^{-\mathrm{i}\pi \alpha_j/2}.
\end{equation}
Notice that at the position of the first vortex $\theta_2=0$ but at the second vortex $\theta_1=\pi$.  This choice of cuts has as a consequence that   functions $A_2(\vec{x}_2)$ and $B_2(\vec{x}_2)$ with  $\vec{x}_2$  centered at the second vortex are given by a slightly different expressions
\begin{equation}
A_2(\vec{x}_2)\underset{|\vec{x}_2|\to\infty }{\longrightarrow} \sqrt{\frac{2}{\pi\mathrm{i} k R}} \mathrm{e}^{\mathrm{i}kR} 
f_2^{(0)}(\theta_2) \mathrm{e}^{\mathrm{i}\alpha_1 (\theta_1-\pi)+\mathrm{i}\alpha_2\theta_2},\qquad
B_2(\vec{x}_2)\underset{|\vec{x}_2|\to\infty }{\longrightarrow} \sqrt{\frac{2}{\pi\mathrm{i}  k R}} \mathrm{e}^{\mathrm{i}kR} 
g_2^{(0)}(\theta_2) \mathrm{e}^{\mathrm{i}\alpha_1 (\theta_1-\pi)+\mathrm{i}\alpha_2\theta_2} . 
\end{equation}
The first order corrections correspond to  re-scattering of these fields on the second vortex.  When coordinates are calculated from the second vortex ($\vec{x}_2=\vec{x}_1-\vec{L} $) and $L\to\infty$,   $A_1(\vec{x}\,)$ has the following asymptotics 
\begin{equation}
\lim_{L\to\infty}A_1(\vec{x}_2)\longrightarrow D(L) f_1^{(0)}(\pi)\mathrm{e}^{\mathrm{i}\alpha_1 \theta_1+\mathrm{i}\alpha_2 \theta_2}  \mathrm{e}^{-\mathrm{i}k x},\qquad 
D(L)=\sqrt{\frac{2}{\pi\mathrm{i} k L}} \mathrm{e}^{\mathrm{i}kL} .
\end{equation}
According to Eq.~\eqref{scattering_function} the scattering function for this incident field is $\mathcal{F}_2(\theta,\pi)$ in Eq.~\eqref{F_pi} without factor  $\mathrm{e}^{\mathrm{i}\alpha_2 \theta}$ which is included in the above definition,
\begin{equation}
\mathcal{F}_2(\theta,\pi)\equiv \mathrm{F}(\theta)\mathrm{e}^{\mathrm{i}\alpha_2 \theta}, \qquad  
\mathrm{F}(\theta)=-\frac{\mathrm{i}\sin \pi \alpha }{2\cos (\theta/2)}\mathrm{e}^{-\mathrm{i}\theta/2}.
\end{equation} 
In \eqref{scattering_function} radius, $R$, is counted from the second vortex. To shift it to the first vortex requires to write $R\approx r+L\cos \theta$.  Therefore the full contribution to function $F_1(\theta, L) $ in two lowest orders is (when $r\to\infty$ $\theta_1=\theta_2=\theta$) 
\begin{equation}
F_1(\theta, L)=\left ( f_1^{(0)}(\theta)+D(L) f_1^{(0)}(\pi )\mathrm{F}_2(\theta)\mathrm{e}^{\mathrm{i}k L\cos \theta}\right )\mathrm{e}^{\mathrm{i}(\alpha_1 +\alpha_2) \theta} .
\end{equation}
In a similar manner
\begin{equation}
G_1(\theta,L)=\left ( g_1^{(0)}(\theta)+D(L) g_1^{(0)}(\pi)\mathrm{F}_2(\theta)\mathrm{e}^{\mathrm{i}k L\cos \theta}\right )\mathrm{e}^{\mathrm{i}(\alpha_1 +\alpha_2) \theta} . 
\end{equation}
Derivatives of functions $F_1(\theta, L)$ and $G_1(\theta, L)$ over  $L$   according to Eqs.~\eqref{diff_L}
are  (in the lowest order)  linear combinations of $F_2^{(0)}(\theta)= f_2^{(0)}(\theta)\mathrm{e}^{\mathrm{i}(\alpha_1 +\alpha_2) \theta} $ and $G_2^{(0)}(\theta)=g_2^{(0)}(\theta)\mathrm{e}^{\mathrm{i}(\alpha_1 +\alpha_2) \theta}$. Performing the calculations one finds that 
\begin{equation}
\left (\begin{array}{cc}\beta_2 & \delta_2\\ \epsilon_2 & \zeta_2 \end{array}\right) \underset{L\to\infty}{\longrightarrow }
\frac{k \sin \pi \alpha_2}{2} 
\left (\begin{array}{cc}   
\dfrac{c_3(\alpha_1)}{ c_3(\alpha_2)} \mathrm{e}^{\mathrm{i}\pi (\alpha_1 -\alpha_2) /2}& 
-\dfrac{\mathrm{i}c_3(\alpha_1)}{c_4(\alpha_2)}\mathrm{e}^{\mathrm{i}\pi(\alpha_1 +\alpha_2)/2}\\
-\dfrac{\mathrm{i}c_4(\alpha_1)}{c_3(\alpha_2)}\mathrm{e}^{-\mathrm{i}\pi (\alpha_1+ \alpha_2)/2}& 
 -\dfrac{c_4(\alpha_1) }{c_4(\alpha_2)}\mathrm{e}^{-\mathrm{i}\pi (\alpha_1 -\alpha_2)/2}
\end{array}\right)  D(L)\mathrm{e}^{\mathrm{i}\pi \alpha_1}\ .
\label{second_L_to_infinity}
\end{equation} 
For the second vortex 
\begin{eqnarray}
F_2(\theta,L)&=& \left( f_2^{(0)}(\theta)+D(L) f_2^{(0)}(0)\mathrm{G}_1(\theta)\mathrm{e}^{-\mathrm{i}k L\cos \theta}\right ) \mathrm{e}^{\mathrm{i}(\alpha_1 +\alpha_2) \theta-\mathrm{i}\pi\alpha_1}\ , \nonumber\\
G_2(\theta,L)&=& \left ( g_2^{(0)}(\theta)+D(L) g_2^{(0)}(0)\mathrm{G}_1(\theta) \mathrm{e}^{-\mathrm{i}k L\cos \theta}\right ) \mathrm{e}^{\mathrm{i}(\alpha_1 +\alpha_2) \theta-\mathrm{i}\pi\alpha_1}\ .
\end{eqnarray}
Here $\mathrm{G}_1(\theta)$ indicates the scattering amplitude $\mathcal{F}_1(\theta,0)$ with factor  $\mathrm{e}^{\mathrm{i}\alpha_1 \theta}$  removed 
\begin{equation}
\mathcal{F}_1(\theta, 0) \equiv \mathrm{G}_1(\theta) \mathrm{e}^{\mathrm{i}\alpha_1\theta} ,\qquad 
\mathrm{G}_1(\theta)=  \frac{\sin \pi\alpha_1}{2\sin (\theta/2)}\mathrm{e}^{-\mathrm{i}\theta/2}.
\end{equation}
 Differentiating them by $L$ and using Eqs.~\eqref{diff_L} one finds
\begin{equation}
\left (\begin{array}{cc}\beta_1 & \delta_1\\ \epsilon_1 & \zeta_1 \end{array}\right) \underset{L\to\infty}{\longrightarrow }
\frac{k \sin \pi \alpha_1}{2} 
\left (\begin{array}{cc}   
-\dfrac{c_3(\alpha_2)}{c_3(\alpha_1)}\mathrm{e}^{-\mathrm{i}\pi (\alpha_1-\alpha_2)/2}&
 -\dfrac{\mathrm{i} c_3(\alpha_2) }{c_4(\alpha_1)}\mathrm{e}^{\mathrm{i}\pi( \alpha_1  +\alpha_2)/2}\\
-\dfrac{\mathrm{i} c_4(\alpha_2)}{ c_3(\alpha_1)} \mathrm{e}^{-\mathrm{i}\pi(\alpha_1 + \alpha_2)/2} &
\dfrac{c_4(\alpha_2)}{ c_4(\alpha_1)} \mathrm{e}^{\mathrm{i}\pi (\alpha_1 - \alpha_2)/2} 
\end{array}\right)D(L)\mathrm{e}^{-\mathrm{i}\pi \alpha_1}\ .
\label{first_L_to_infinity}
\end{equation}
These asymptotic values obey Eq.~\eqref{relation_2_1} with 
\begin{equation}
\rho= \frac{\alpha_1(1-\alpha_1)}{\alpha_2(1-\alpha_2)}\mathrm{e}^{2\pi \mathrm{i}\alpha_1}
\end{equation} 
in agreement with \eqref{t_one}.    

The asymptotic behavior of variables \eqref{variables} is
\begin{equation}
y\equiv \delta_2\epsilon_1 \underset{L\to\infty }{\longrightarrow} -\tfrac{1}{4}k^2\sin \pi \alpha_1 \sin \pi\alpha_2 D^2(L)\ ,\qquad 
z\equiv \beta_1\beta_2 \underset{L\to\infty }{\longrightarrow} -\tfrac{1}{4}k^2\sin \pi \alpha_1 \sin \pi\alpha_2 D^2(L)\ .
\label{additional_L_to_infinity}
\end{equation}
\section{Conclusion}\label{conclusion}

The main result of the paper is the demonstration that the  problem of two Aharonov-Bohm vortices is integrable. As it is often in integrable systems, the exact solution is lengthy and tedious. 

The solution has been obtained by a generalization of the method used in Ref.~\cite{myers} to solve scalar  diffraction problem on scattering on a finite slit in 2 dimensions. The principal steps of the solution are the following.
\begin{itemize}
\item Due to singular nature of AB interactions, wave functions are fixed uniquely by their behavior in small vicinity of the vortices and only a finite number of coefficients is necessary to reconstruct wave functions. 
\item  To find these coefficients it is useful to introduce singular functions $A_j(\vec{x}\,)$ and $B_j(\vec{x}\, )$ independent on incident fields with prescribed singularities at vortex $j$ (see Eqs.~\eqref{A_j} and \eqref{B_j}). 
\item For the Helmholtz equation in the plane (and in other symmetric space as well) there exists a group of first order differential  transformations which commute with the Lagrangian and cancels the incident field. 
\item  When any of such transformations is applied to the exact wave function, the resulting function corresponds to zero incident field. But, in general, the transformed function becomes singular in one or many vortices. As all invariant operators are of the first order, these singularities can be compensated by a suitable linear combination of auxiliary singular functions $A_j(\vec{x}\,)$ and $B_j(\vec{x}\, )$. In such a manner one gets a large set of equations which express  certain derivatives of the Green function and the scattering amplitude through functions $A_j(\vec{x}\,)$ and $B_j(\vec{x}\, )$ (see Eqs.~\eqref{partial_Lj_two_vortices}, 
\eqref{dy_dy_prime}, \eqref{rotation_G},  \eqref{f_g}).   
\item Specializing these relations to vicinity of vortex positions proves that certain derivatives of  functions  $A_j(\vec{x}\, )$ and $B_j(\vec{x}\, )$ are linear combinations of the same functions (see Eqs.~\eqref{dif_A_B_two_vortices}, \eqref{translation_1}, \eqref{translation_2},  \eqref{rotation_2}).  
\item Coefficients in these relations are functions of vortex separations and by calculating commutators of different group transformations one obtains a system of non-linear equations for them (see Eqs.~\eqref{eqs_different_fluxes}-\eqref{y_beta_u_w}).   
\item All necessary coefficients can be calculated from a solution of the Painlev\'e V  equation \eqref{main_equation} or (after a non-linear B\"acklund transformation)  of the Painlev\'e III equation  \eqref{w_equation}.
\item As all equations are differential, to really use them  it is necessary to know values of coefficients in a certain point. Analytically, one can calculate  asymptotics of these coefficients in the limit  $L\to 0$   (see Eqs.~\eqref{second_limit}-\eqref{L_to_zero}, \eqref{first_L_to_zero_1}-\eqref{second_L_to_zero_1},\eqref{additional_L_to_zero_0}-\eqref{y_L_to_zero_0}) and/or  $L\to \infty$ (see Eqs.~\eqref{second_L_to_infinity}-\eqref{additional_L_to_infinity}). 
\end{itemize}

The method of Ref.~\cite{myers} used throughout the paper is quite general and flexible. Originally it has  been  used for solving certain integral equations, see Refs.~\cite{myers} and \cite{atas}. 
As it is demonstrated in this paper, it can also be  adapted to the  problem  of scattering on two AB vortices. Its generalizations for similar problems for  the Klein-Gordon and Dirac operators in the Minkowski and Euclidean spaces (and probably in other symmetry spaces as well) seem to be possible.


\subsection*{Notes added}

The principal ingredient of the above solution was the adaptation of the method of Ref.~\cite{myers} to problems of singular AB vortices. After the paper has been practically finished, O. Lisovyy has remarked to the author that similar equations (even for an arbitrary number of vortices) had been derived by M. Sato, T. Miwa, and M. Jimbo in Ref.~\cite{swj} in a different manner. That work is one in the long  series of papers  devoted to developments of the theory of holonomic quantum fields (see e.g. \cite{jimbo} and references therein). In Ref.~\cite{swj}, the authors constructed wave function with prescribed monodromy around a finite number of points. In two dimensions, monodromy transformations for the scalar equation reduce to the appearance of the phase factor $\mathrm{e}^{2\pi \mathrm{i}\alpha_j}$ after encircling a point $j$ which  corresponds exactly to a AB flux line at this point. To get the necessary equations, the authors of  Ref.~\cite{swj} wrote the most general behavior of auxiliary functions  $A_j(\vec{x}\,)$ and $B_j(\vec{x}\, )$ in small vicinity of vortex positions as in Eqs.~\eqref{expansions_AB_1}, \eqref{expansions_AB_2}. Computing the action of operators commuting with  the Lagrangian and using the uniqueness as has been done in previous Sections one gets the same system of equations as above. 

The main differences of this paper and of Ref.~\cite{swj} is in the later the Euclidean space has been considered. So the Klein-Gordon equation (i.e. the Helmholtz equation \eqref{helmholtz} with reversed sign of $k^2$) is 
\begin{equation}
(\partial_x^2+\partial_y^2-k^2)\Psi(\vec{x}\,)=0.
\end{equation}
The analogue of the radiation condition \eqref{radiation} in this case takes the form
\begin{equation}
 \Psi(\vec{x}\,)\underset{R\to \infty}{\longrightarrow} \frac{1}{\sqrt{R}}\mathrm{e}^{-kR}F(\theta).
\end{equation}
Therefore, complex conjugation of the solution does not change the correct asymptotics at infinity and wave function with all opposite fluxes (necessary in Section \ref{determination_constants}) is simply the complex conjugate of the initial wave function. In the Minkowski space used throughout the paper, complex conjugate turns out-going waves to in-going ones and is not an allowed transformation.  For two vortices in Minkowski space these two functions are related by the inversion at the line connecting the vortices (see Eq.~\eqref{S_symmetry}). In general, wave functions with opposite fluxes appeared in the reciprocity relation should be calculated separately which effectively  double  the number of unknown variables.

Another difference between this  paper and  Ref.~\cite{swj} is that in the latter the question of correct limiting values of  necessary variables has not been discussed. Even for two vortices calculations of wave functions in the limit of small and large vortex separation is a complicated problem (see Section~\ref{small_distance}). For larger number of vortices it remains an open question.   

In general, interrelations of the theory of holonomic quantum fields and the AB problem is not widely known and fully understood (e.g. there is no reference in \cite{swj} to the  paper of Aharonov and Bohm, Ref.~\cite{AB}) and further investigation of this subject is of interest.      
    
\acknowledgments

The author is greatly indebted to Oleg Lisovyy for explication  the relations between theory of holonomic quantum fields, especially Ref.~\cite{swj}, and the problem of a few AB vortices and to St\'ephane Ouvry for many stimulating discussions.


\appendix


\section{Uniqueness of solution and reciprocity relation for scattering on AB vortices}\label{uniqueness}

The standard way of proving general statements about wave equation solutions is the use of current conservation. Let $\Psi_1$ and $\Psi_2$ be two solutions of the Helmholtz equation \eqref{helmholtz}. The current conservation means that   
\begin{equation}
\oint \vec{J}\mathrm{d}\vec{s}=0
\label{conservation}
\end{equation}
where  current $\vec{J}$ is 
\begin{equation}
\vec{J}= \Psi_2 \partial_{\vec{x}} \Psi_1 -\Psi_1 \partial_{\vec{x}}  \Psi_2,
\end{equation}
and the integration is performed along any closed contour inside which there is no singularities of $\Psi_{1,2}$. For two AB vortices  possible  contours of integration can be chosen  as in Fig.~\ref{contour}.

To cancel the current along the both sides of cuts due to phase jumps it is necessary to choose boundary jumps of $\Psi_1$ and $\Psi_2$ differently. If   $\Psi_1$ obey conditions \eqref{boundary} then to have zero current through the cuts function  $\Psi_2$ should obey the same conditions but with reversed signs of all fluxes
\begin{equation}
\Psi_{2+}(x,0)=\mathrm{e}^{-2\pi \mathrm{i}\alpha_j}\Psi_{2-}(x,0),\qquad 
\partial_y \Psi_{2+}(x,0)=\mathrm{e}^{-2\pi \mathrm{i}\alpha_j}\partial_y \Psi_{2-}(x,0),\qquad x\in \mathcal{C}_j.
\end{equation}
When these conditions are fulfilled, it remains to check conservation of current along two other types of contours. The first  consists of small circles around each vortex. If the both functions obey regularity condition \eqref{asymptotic_j},  the integral of the current over such  circles  tends to zero with decreasing of the radius. The last contour is the circle of  large radius encircling all vortices (cf. Fig.~\ref{contour}). Its treatment depends on the problem considered. 

If there exist two solutions corresponding to scattering on AB vortices  then their difference, $\delta \Psi$, obey the Helmholtz equation and all conditions  \eqref{boundary}-\eqref{asymptotic_j}  with zero incident field. Choosing $\Psi_1=\delta \Psi$ and $\Psi_2=\delta \Psi^*$ one concludes that the conservation of current implies that  
 \begin{equation}
\lim_{R\to\infty}R \int_0^{2\pi }(\delta \Psi^* \partial_R \delta \Psi -\delta \Psi \partial_R \delta \Psi ) \mathrm{d}\phi=0.
\end{equation}
But 
\begin{equation}
|\partial_R \delta \Psi -\mathrm{i}  k \delta \Psi|^2=|\partial_R \delta\Psi|^2+k^2 |\delta \Psi|^2 +\mathrm{i}  k\Big (  \delta \Psi^* \partial_R \delta \Psi -\delta \Psi \partial_R \delta \Psi \Big)\ .
\end{equation}
Using radiation condition \eqref{radiation} and the previous expression one concludes that if $k\neq 0$
\begin{equation}
\lim_{R\to\infty}R\int_0^{2\pi}|\delta \Psi|^2\mathrm{d}\phi=0,\qquad \lim_{R\to\infty}R\int_0^{2\pi}|\partial_R \delta\Psi|^2\mathrm{d}\phi=0.
\label{limits}
\end{equation}  
Let $\nu=\sum_j{\alpha_j}$ be the total flux of the vertices.  Outside the circle of radius $R$ which includes all vortices  function $\delta \Psi$ can be expanded in formal series on Hankel functions 
\begin{equation}
\delta \Psi =\sum_{n=-\infty}^{\infty} A_n H_{\nu+n}^{(1)}(kr)\mathrm{e}^{\mathrm{i}(\nu+n)\phi }\ .
\end{equation}
As Hankel functions decrease when $r\to\infty$ as $r^{-1/2}$, from \eqref{limits} it follows that all $|A_n|^2 =0$. As these coefficients are coefficients of expansion over a complete set of functions one concludes that the only possibility is that $\delta \Psi\equiv 0$. In other words, the only solution obeying all conditions \eqref{boundary}-\eqref{asymptotic_j}  with zero incident field is identically zero.   
\vspace{.4cm}

Similar arguments are used to find the reciprocity relation which  relates the Green functions with interchanged positions of the source point and the observation point. For the scalar diffraction problems these two functions are equal but for scattering on AB-vortices one has to reverse all vortex fluxes. Let us denote the Green function for the scattering on vortices $\vec{\alpha}=\alpha_1,\ldots,\alpha_n$ by $G_{\vec{\alpha}}(\vec{x},\vec{x}^{\, \prime})$ ($0<\alpha_j<1$). Then the reciprocity relation reads
\begin{equation}
G_{\vec{\alpha}}(\vec{x},\vec{x}^{\, \prime})=G_{-\vec{\alpha}}(\vec{x}^{\, \prime},\vec{x}\,) 
\label{recip} 
\end{equation}
where $-\vec{\alpha}=1-\alpha_1,\ldots,1-\alpha_n$. 

The proof of this formula can be done as follows. 
By definition, the both functions obey the Helmholtz equation
\begin{equation}
(\Delta +k^2)G_{\vec{\alpha}}(\vec{x},\vec{x}^{\, \prime}) =\delta(\vec{x}-\vec{x}^{\, \prime}),\qquad 
(\Delta +k^2)G_{-\vec{\alpha}}(\vec{x},\vec{x}^{\, \prime \prime}) =\delta(\vec{x}-\vec{x}^{\, \prime \prime}).
\label{psi_alpha}
\end{equation}
and on any cut $\mathcal{C}_j$ they have opposite phase jumps. 

From Eqs.~\eqref{psi_alpha} it follows that
\begin{equation}
G_{\vec{\alpha}}(\vec{x}^{\, \prime\prime},\vec{x}^{\, \prime})-G_{-\vec{\alpha}}(\vec{x}^{\, \prime},\vec{x}^{\, \prime\prime })=
\oint \vec{J}\mathrm{d}\vec{s}
\end{equation}
where 
\begin{equation}
\vec{J}(\vec{x}\,)=G_{\vec{\alpha}}(\vec{x},\vec{x}^{\, \prime}) \partial_{\vec{x}}\, G_{-\vec{\alpha}} (\vec{x},\vec{x}^{\, \prime \prime})- G_{-\vec{\alpha}}(\vec{x},\vec{x}^{\, \prime \prime}) \partial_{\vec{x}}\, G_{\vec{\alpha}}(\vec{x},\vec{x}^{\, \prime}) ,
\end{equation} 
with the integration is being taken over the same contour as above. As the both Green functions obey the same radiation conditions \eqref{radiation} the current over a big circle tends to zero which proves reciprocity relation \eqref{recip}. 
\vspace{.4cm}

In \cite{swj} the Euclidean case, $k^2<0$, has been considered. As a consequence,   
$\Psi_{-\vec{\alpha}}(\vec{x}\, )=\Psi_{\vec{\alpha}}^*(\vec{x}\, ) $, as in this case  the complex conjugation does not change the asymptotic of wave function on infinity ($\Psi\sim \mathrm{e}^{-kr}$). For real $k$ when $\Psi\sim \mathrm{e}^{\mathrm{i}kr}$ complex conjugation contradicts  the radiation condition \eqref{radiation} and  $\Psi_{-\vec{\alpha}}(\vec{x}\, )$ is not related immediately with  $\Psi_{\vec{\alpha}}(\vec{x}\, )$. For the problem of two vortices there exists an additional symmetry, namely the reflection in the line connecting two vortices. As such inversion interchange upper and lower parts of the cuts, conditions \eqref{boundary} are now fulfilled but with opposite fluxes  (in other words,   fluxes are pseudo-scalars). Another method to check this relation is to consider small-$x$ behavior of wave functions  Eqs.~\eqref{asymptotic_j}. It is clear that the inversion of  $y$-coordinate is equivalent to reversing the flux,  $\alpha\to 1-\alpha$. Therefore up to a phase factor
\begin{equation}
\Psi_{\vec{\alpha}}(\vec{x}\, )=\Psi_{-\vec{\alpha}}(\hat{S}\vec{x}\, )
\label{S_symmetry}
\end{equation}
where the inversion $\hat{S}$ acts as follows
\begin{equation}
\hat{S}(x,y)=(x,-y).
\end{equation}
Combining it with Eq.~\eqref{recip}, one concludes that the following form of the reciprocity is valid for the two vortex problem
\begin{equation}
G_{\vec{\alpha}}(\vec{x},\vec{x}^{\, \prime})=G_{\vec{\alpha}}(\hat{S}\vec{x}^{\, \prime},\hat{S}\vec{x}\,) .
\label{full_reciprocity}
\end{equation}
To determine constants $t_j$  in \eqref{A_a} one can proceed as follows. Close to a vortex with flux $\alpha$ situated at $0$  the Green function behaves as in \eqref{x_Lj_alpha_j}
\begin{equation}
G(\vec{x} ,\vec{x}^{\, \prime}) \underset{x\to 0}{\longrightarrow} 
a(\vec{x}^{\, \prime}) \, (x+ \mathrm{i}y)^{\alpha}+ 
b(\vec{x}^{\, \prime}) \, (x -\mathrm{i}y)^{1-\alpha} .
\label{x_to_zero}
\end{equation}
Assume for simplicity that the cut associated with this vortex lies on negative $x$-axis. When the position of the vortex is shifted from $0$ to $\delta L$ along the $x$-axis the new field close to the vortex remains practically unchanged
\begin{equation}
G_{\delta L }(\vec{x} ,\vec{x}^{\, \prime}) \underset{x \to \delta L}{\rightsquigarrow} 
a(\vec{x}^{\, \prime}) \, (x-\delta L+ \mathrm{i}y)^{\alpha}+ 
b(\vec{x}^{\, \prime}) \, (x-\delta L -\mathrm{i}y)^{1-\alpha} 
\label{new_field}
\end{equation}
but a new portion of the cut from $0$ to $\delta L$ appears. (Implicitly it is assumed that $\delta L>0$ so the length of the cut increases). From \eqref{new_field} it follows that the difference of the field on the both side of such new cut is (for $x\in [0,\delta L] $)
\begin{eqnarray}
\left [G^{(+)}_{\delta L }(\vec{x} ,\vec{x}^{\, \prime})-G^{(-)}_{\delta L}(\vec{x} ,\vec{x}^{\, \prime})\right ]_{y=0}&=& 2\mathrm{i}\sin \pi \alpha \left [
a(\vec{x}^{\, \prime})\nu_1(x)-b(\vec{x}^{\, \prime} ) \nu_2(x)\right ],\\
\left [\partial_y G^{(+)}_{\delta L } (\vec{x} ,\vec{x}^{\, \prime})-\partial_y G^{(-)}_{\delta L }(\vec{x} ,\vec{x}^{\, \prime})\right ]_{y=0}&=&
2\mathrm{i}\sin \pi \alpha \left [
a(\vec{x}^{\, \prime})\, \mathrm{i}\nu_1^{\prime }(x)+b(\vec{x}^{\, \prime})\, \mathrm{i} \nu_2^{\prime }(x)\right ]
\end{eqnarray} 
where 
\begin{equation}
\nu_1(x)=(\delta L-x)^{\alpha},\qquad \nu_2(x)=(\delta L-x)^{1-\alpha}.
\end{equation}
These expressions determine the field on the both sides of a cut $[0,\delta L]$. According to the Green theorem the  field everywhere is given by the integral
\begin{eqnarray}
G_{\delta L }(\vec{x} ,\vec{x}^{\, \prime})&=& G(\vec{x} ,\vec{x}^{\, \prime})+  \int_{0}^{\delta L} G(\vec{x} ,\vec{z}\, )\left [\partial_y G^{(+)}_{\delta L } (\vec{x} ,\vec{x}^{\, \prime})-\partial_y G^{(-)}_{\delta L }(\vec{x} ,\vec{x}^{\, \prime})\right ]_{y=0}\mathrm{d}z \nonumber \\
&-&
\int_{0}^{\delta L}\partial_{r} G(\vec{x} ,\vec{z}\, )\left [G^{(+)}_{\delta L }(\vec{x} ,\vec{x}^{\, \prime})-G^{(-)}_{\delta L}(\vec{x} ,\vec{x}^{\, \prime})\right ]_{y=0}\mathrm{d}z \ .
\end{eqnarray}
Here $\vec{z}$ denotes  point $(z,r)$ with $r\to 0$. 

Due to the reciprocity relation \eqref{full_reciprocity} the behavior of the Green function $G(\vec{x} ,\vec{x}^{\, \prime}) $ when $x^{\prime}\to 0$ is the follows (cf. \eqref{x_to_zero}) 
\begin{equation}
G(\vec{x} ,\vec{x}^{\, \prime}) \underset{x^{\prime}\to 0}{\longrightarrow} 
a(S\vec{x}\, ) \, (x^{\, \prime}- \mathrm{i}y^{\, \prime})^{\alpha}+ 
b(S\vec{x}\,) \, (x^{\, \prime} +\mathrm{i}y^{\, \prime})^{1-\alpha} ,\qquad x^{\, \prime}=z,\; y^{\, \prime}=r\to 0.
\end{equation}
Therefore in the leading order in $\delta L$ the Green function is
\begin{equation}
G_{\delta L }(\vec{x} ,\vec{x}^{\, \prime})\approx  G(\vec{x} ,\vec{x}^{\, \prime})
-2\sin \pi \alpha\   a(S\vec{x}\, ) [a(\vec{x}^{\, \prime}) q_1+ b(\vec{x}^{\, \prime}  )r_1]  
-2\sin \pi \alpha\   b(S\vec{x} \, )[b(\vec{x}^{\, \prime} )q_2+a(\vec{x}^{\, \prime} )r_2]  
\end{equation} 
where
\begin{eqnarray}
q_1&=&   \int_0^{\delta L}[ z^{\alpha } \nu_1^{\prime}(z)+  \alpha z^{\alpha-1} \nu_1(z)]\mathrm{d}z,\qquad
r_1=  \int_0^{\delta L}[ z^{\alpha } \nu_2^{\prime}(z) -\alpha z^{\alpha-1}\nu_2 (z)]\mathrm{d}z, \\
q_2&= & \int_0^{\delta L}[ z^{1-\alpha } \nu_2^{\prime}(z) +(1-\alpha) z^{-\alpha}\nu_2 (z)]\mathrm{d}z, \qquad 
r_2=   \int_0^{\delta L}[ z^{1-\alpha } \nu_1^{\prime}(z)- (1- \alpha) z^{-\alpha} \nu_1(z)]\mathrm{d}z.
\end{eqnarray}
It is plain that $q_j=0$ and 
\begin{equation}
r_1=r_2=-\frac{2\pi \alpha(1-\alpha)}{\sin \pi \alpha}\delta L.
\end{equation}
It means that one gets \eqref{A_a} with 
\begin{equation}
t_j=4\alpha_j(1-\alpha_j).
\end{equation}
The same result follows from the one-vortex solution (see Appendix~\ref{one_vortex}).


\section{One-vortex solution and  local constants}\label{one_vortex}

The Green function for the AB problem with one vortex with flux $\alpha$ is 
\begin{equation}
G(\vec{x},\vec{x}^{\, \prime})=\frac{1}{4\mathrm{i}}\sum_{n=-\infty}^{\infty}\left \{ \begin{array}{cc} 
J_{|n+\alpha|}(kr)\,  \mathrm{e}^{\mathrm{i}(n+\alpha)\theta }\,  H_{|n+\alpha|}^{(1)}(kR)\, \mathrm{e}^{-\mathrm{i}(n+\alpha)\phi}  ,& r<R\\
J_{|n+\alpha|}(kR) \, \mathrm{e}^{-\mathrm{i}(n+\alpha)\phi} \,  H_{|n+\alpha|}^{(1)}(kr)\, \mathrm{e}^{\mathrm{i}(n+\alpha)\theta},& r>R
\end{array}\right . \ .
\label{one_vortex_G}
\end{equation}
Here $\vec{x}=(r\cos \theta, r\sin \theta)$ and $\vec{x}^{\, \prime}=(R\cos \phi, R\sin \phi)$. 

When $0<\alpha<1$,  terms with $n=0$ and $n=-1$ dominate  in the limit  $|\vec{x}\, |\to 0$ and as \cite{bateman}
\begin{equation}
J_{\nu}(r)\underset{r\to 0}{\longrightarrow}\left(\frac{r}{2}\right )^{\nu}\frac{1}{\Gamma(1+\nu)}
\end{equation}
it follows that 
\begin{equation}
G(\vec{x},\vec{x}^{\, \prime})\underset{|\vec{x}\, |\to 0}{\longrightarrow} a(\vec{x}^{\, \prime})(x+\mathrm{i}y)^{\alpha}+ 
b(\vec{x}^{\, \prime} )(x-\mathrm{i}y)^{1-\alpha}\ .
\label{one_vortex_asymptotic}
\end{equation}
Here 
\begin{equation}
a(\vec{x}^{\, \prime})=  c_1(\alpha) H_{\alpha}^{(1)}(kR)\mathrm{e}^{-\mathrm{i}\alpha \phi},\qquad  b(\vec{x}^{\, \prime})= c_2(\alpha) H_{1-\alpha}^{(1)}(kR)\mathrm{e}^{-\mathrm{i}(\alpha-1) \phi}, 
\end{equation}
with 
\begin{equation}
 c_1(\alpha)= -\frac{\mathrm{i}\, k^{\alpha}}{  2^{\alpha+2}  \Gamma(1+\alpha)}\ , \qquad  c_2(\alpha)=-\frac{\mathrm{i}\, k^{1-\alpha}}{  2^{3-\alpha}  \Gamma(2-\alpha)} \ .
 \label{c_1_2}
\end{equation}
From recursive relations for  an arbitrary Bessel function $Z_{\nu}$ and  the definition of the Hankel function  \cite{bateman}
\begin{equation}
Z_{\nu}^{\prime}(r)\pm \frac{\nu}{r}Z_{\nu}(r)=\pm Z_{\nu\mp 1}(r),\quad 
H_{\nu}^{(1)}=\frac{1}{\mathrm{i}\sin \pi \nu}\left(J_{-\nu} -\mathrm{e}^{-\mathrm{i}\pi\nu} J_{\nu}\right),\quad 
H_{-\nu}^{(1)}=\mathrm{e}^{\mathrm{i}\pi\nu}H_{\nu}^{(1)}
\end{equation}
it is straightforward to check that
\begin{equation}
\partial_L G(\vec{x},\vec{x}^{\, \prime})=-\frac{k \sin \pi \alpha}{8} \left [H_{\alpha}^{(1)}(kr) \mathrm{e}^{\mathrm{i}\alpha \theta} H_{1-\alpha}^{(1)}(kR) \mathrm{e}^{\mathrm{i}(1-\alpha) \phi}+ 
H_{1-\alpha}^{(1)}(kr) \mathrm{e}^{-\mathrm{i}(1-\alpha) \theta} H_{\alpha}^{(1)}(kR) \mathrm{e}^{-\mathrm{i}\alpha \phi}    \right ]\ .
\end{equation}
Eq.~\eqref{partial_Lj_two_vortices} states that the derivative over vortex position has the form
\begin{equation}
\frac{\partial G(\vec{x} ,\vec{x}^{\, \prime})}{\partial L}=a(\vec{x}^{\, \prime}) A(\vec{x}\, )+b(\vec{x}^{\, \prime}) B(\vec{x}\,).
\end{equation}
As expected, for the one-vortex problem  functions $A(\vec{x}\, )$ and $B(\vec{x}\, )$  are proportional to the Hankel functions 
\begin{equation}
 A(\vec{x}\, )= c_3(\alpha)  H_{1-\alpha}^{(1)}(kr) \mathrm{e}^{\mathrm{i}(\alpha-1) \theta}  ,\qquad B(\vec{x}\, )= c_4(\alpha)  H_{\alpha}^{(1)}(kr) \mathrm{e}^{\mathrm{i}\alpha  \theta},
 \label{A_B_one} 
 \end{equation}
 with
 \begin{equation}
 c_3(\alpha)=-\mathrm{i} \sin \pi \alpha\,  k^{1-\alpha}2^{\alpha-1}\Gamma(1+\alpha) ,\qquad 
  c_4(\alpha)=-\mathrm{i} \sin \pi \alpha\,  k^{\alpha}2^{-\alpha}\Gamma(2-\alpha) .
\label{c_3_4}  
\end{equation} 
The above expressions permit to calculate local constants for well-separated vortices.  For the one-vortex solution one obtains 
\begin{equation}
(\partial_x+\mathrm{i} \partial_y)A= g_2 B,\qquad g_2=k\mathrm{e}^{\mathrm{i}\alpha \pi }\frac{c_3}{c_4}, \qquad 
(\partial_x-\mathrm{i} \partial_y)B= g_3 A ,\qquad g_3=-k\mathrm{e}^{-\mathrm{i}\alpha  \pi }\frac{c_4}{c_3}.
\end{equation}
From these relation we conclude that diagonal terms $g_1=0$ and $g_4=0$. As these values are independent on $L$, they  always remain zero.
 
From Eqs.~\eqref{rotation_final} one finds that
\begin{equation}
m_1=1-\alpha_1,\qquad n_1= -\alpha_1,\qquad m_2=1-\alpha_2,\qquad n_2=-\alpha_2.
\label{m_n}
\end{equation}
These constants are independent on $L$ and it is these values that are used in the main text. 

According to the reciprocity relation \eqref{A_a} one should have
\begin{equation}
 A(\vec{x}\, )=t\,  b(\hat{S}\vec{x}\,),\qquad B(\vec{x}\, )=t\,  a(\hat{S}\vec{x}\,).
\end{equation}
Here transformation $\hat{S}$ changes sigh of the second coordinate,  $ \hat{S}(x,y)=(x,-y)$. Its explicit form in polar coordinates depends on the choice of cut direction. $\hat{S}(r,\theta)=(r,2\pi \xi-\theta)$ where $\xi=0$ if the cut is along negative $x$-axis, i.e. $-\pi <\theta <\pi$, and $\xi=1$  if the cut is chosen along positive $x$-axis, i.e.  $0<\theta<2\pi$.  Comparing with the above formulas gives $|t|=c_3(\alpha)/c_2(\alpha)=4\pi \alpha(1-\alpha)$ and 
\begin{equation}
t=4\pi \alpha(1-\alpha)\mathrm{e}^{-2\pi \alpha \mathrm{i}\xi}.
\label{t_one}
\end{equation}
From the asymptotics of Eqs.~\eqref{A_B_one} it follows that for the one-vortex solution functions $F(\theta)$ and $G(\theta)$ in Eqs.~\eqref{F_j_G_j} and \eqref{f_g} are
\begin{equation}
F(\theta)=c_3(\alpha) \mathrm{e}^{\mathrm{i} (\alpha-1)\theta -\mathrm{i}\pi (1-\alpha)/2},\qquad G(\theta)=c_4(\alpha) \mathrm{e}^{\mathrm{i} \alpha \theta -\mathrm{i}\pi \alpha /2}.
\label{zeroth}
\end{equation}  
Using \eqref{f_g} and the above values of $c_j(\alpha)$ one finds that the scattering amplitude for the one-vortex problem has the form
(it is assumed that the cut is along the negative $x$-axis, i.e. $-\pi\leq \theta,\phi\leq \pi$)
\begin{equation}
\mathcal{F}(\theta,\phi)=\frac{\sin \pi \alpha}{2\sin \left((\theta-\phi)/2 \right)}\mathrm{e}^{\mathrm{i}(\alpha-1/2)(\theta-\phi) +\mathrm{i}\pi \alpha\,  \mathrm{sign}\, \phi}\ .
\label{f_ab}
\end{equation}  
In particular for $\phi=\pm \pi$ the scattering amplitude is   
\begin{equation}
\mathcal{F}(\theta,\pm \pi)=-\mathrm{i}\sin \pi \alpha \frac{\mathrm{e}^{\mathrm{i}\alpha \theta-\mathrm{i}\theta/2}}{2\cos (\theta/2)}
\label{F_pi}
\end{equation}
which agrees with \cite{AB} (with a correction \cite{hagen}). 


\end{document}